\documentclass[12pt,twocolumn]{aastex62}
\usepackage{lineno}

\usepackage{hyperref}
\usepackage{amsmath,mathtools,mathrsfs,amssymb}
\usepackage{graphicx}
\usepackage{bm}
\usepackage[percent]{overpic}

\defcitealias{Medina-Torrejon_2021}{MGK+21}
\defcitealias{Medina-Torrejon_2023}{MGK23}
\defcitealias{xu_lazarian_2023}{XL23}

\submitjournal{ApJ}

\usepackage{multirow}

\begin{document}

\title{Particle Acceleration Time due to Turbulence-Induced Magnetic Reconnection}
\email{dalpino@iag.usp.br,temt@usp.br}

\author[0000-0001-8058-4752]{Elisabete M. de Gouveia Dal Pino}
\affiliation{Universidade de S\~{a}o Paulo, Instituto de Astronomia, Geof\'{i}sica e Ci\^{e}ncias Atmosf\'{e}ricas, Departamento de Astronomia, 1226 Mat\~{a}o Street, S\~{a}o Paulo, 05508-090, Brazil}

\author[0000-0003-4666-1843]{Tania E. Medina-Torrej\'{o}n}
\affiliation{Universidade de S\~{a}o Paulo, Instituto de F\'{i}sica de São Carlos, S\~{a}o Paulo, Brazil}




\begin{abstract}
In this work, we numerically investigate a crucial parameter in turbulence-induced magnetic reconnection theory: the particle acceleration time. Employing simulations of magnetically dominated turbulent relativistic jets, we examine particles accelerated either during the jet's evolution or in a nearly stationary post-processed state. We derive acceleration times and compare them with theoretical predictions for both the Fermi and drift regimes identified in the simulations.
In the Fermi regime, theory predicts that acceleration time remains nearly constant with particle energy for a fixed reconnection velocity, as energy grows exponentially in time. Conversely, we expect the reconnection acceleration time to depend on the current sheet's thickness and the reconnection velocity, a dependence recently revisited by \citet{xu_lazarian_2023}. We test their relations using statistical distributions of current sheet thicknesses, reconnection speeds  and find the average acceleration time agrees well with their predictions, showing weak dependence on particle energy.
We further validate these findings by comparing them with in situ acceleration times from 50,000 test particles. The results show 
good agreement, particularly for the fastest reconnection acceleration conditions.  When simulating longer acceleration periods in a quasi-steady turbulent jet snapshot, the acceleration time during the Fermi regime remains nearly constant up to a critical energy—where the particles’ Larmor radius matches the largest current sheet thickness. Above this threshold, particles enter the slower drift regime, where acceleration time becomes strongly energy-dependent, as expected. Overall, our results confirm the theoretical framework and highlight the dominance of the Fermi acceleration process up to very high energies.

\end{abstract}



\section{Introduction} \label{sec:intro}








Recent advancements in high-energy astrophysics have  underscored the crucial role of magnetic reconnection in accelerating energetic particles beyond the framework of solar system. Magnetic reconnection is now recognized as  a strong candidate in the generation of very and  ultra-high-energy cosmic rays (UHECRs) as well as the very high-energy (VHE) flares produced by them within magnetically dominated regions of  sources like accretion flows and relativistic jets around black holes and compact sources, pulsar wind nebulae, and GRBs 
\citep[e.g.][]{dalpino_lazarian_2005,giannios_etal_09,dgdp_etal_10, dalpino_etal_2010b, zhang_yan_11, mckinney_2012,arons2013,kadowaki_etal_15, singh_etal_15,zhangli2015,zhang_etal_2018, cerutti_etal_2013, yuan_etal_2016, lyutikov_etal_2018, petropoulou_etal_2016, christie_etal_19, Medina-Torrejon_2021, Murase2022, Medina-Torrejon_2023, zhang2021, Zhang2023}. 

The understanding of particle acceleration driven by magnetic reconnection has significantly advanced, benefiting from both particle-in-cell (PIC) simulations, primarily conducted in two-dimensional (2D) settings
\citep[e.g.,][]{zenitani_H_2001,drake_etal_2006,zenitani_H_2007,zenitani_H_2008,lyubarsky_etal_2008,drake_etal_2010,clausen-brown_2012,cerutti_etal_2012,cerutti_etal_2014,li_etal_2015,werner_etal_2018,werner_etal_2019,lyutikov_etal_2017,sironi_spitkovsky_2014,guo_etal_2015,guo_etal_2016,guo_etal_2020,sironi_etal_2015,ball_etal_2018,
comisso18, guo_etal_2019,kilian_etal_2020, comisso_sironi_2021, zhang2021, Zhang2023, Guo2023},
and magnetohydrodynamic (MHD) simulations, mainly performed in three dimensions (3D)
\citep[e.g.,][]{kowal_etal_2011,kowal_etal_2012, dalpino_kowal_15, lazarian12, delvalle_etal_16, beresnyak_etal_2016, Ripperda2017, kadowaki_etal_2021, Medina-Torrejon_2021,  Medina-Torrejon_2023}.



 The main results found from these studies can be succinctly summarized as follows.
 
PIC simulations  explore  scales of the order of 100-1000 times the plasma inertial length ($c/\omega_{\rm p}$, where $c$ is the light speed and $\omega_{\rm p}$ is the plasma frequency), representing microscopic scales several orders of magnitude smaller than astrophysical source scales (typically, around 
$10^{-10}$ and $10^{-17}$ 
times smaller than solar and relativistic jet scales, respectively).
In these simulations, fast reconnection is primarily driven by the tearing mode instability. In two-dimensions  this leads to the formation of plasmoids (magnetic islands) which are actually 
 confined to 2D space geometry \citep[][]{Vicentin2024}.
The reconnection velocity ($v_{\rm rec}$) in this case depends on the plasma resistivity ($\eta$) and 
particle acceleration can reach energies of up to a few $10^2$ times the particle rest mass energy ($mc^2$) only.
The dominant electric field responsible for particle acceleration is resistive,  associated with the current density ($\eta \mathbf{J}$).

On the other hand, magnetohydrodynamic (MHD) simulations with the injection of test particles, are designed to explore the macroscopic astrophysical scales of the process. In this regime,
fast reconnection is predominantly driven by the ubiquitous 3D turbulence in these environments, as a direct consequence of magnetic flux violation 
\citep[][]{lazarian_vishiniac_99, kowal_etal_09, eyink2013, takamoto_etal_15, dalpino_kowal_15, santoslima_etal_2010, Jafari_2018, santos-lima_etal20, Lazarian2020review,  Vicentin2024}.
 Turbulence can be induced by instabilities like Kelvin-Helmholtz
\citep[][]{kowal_etal_2020},
magnetorotational (MRI) \citep[][]{Kadowaki_2018},
current-driven kink (CDKI) \citep[][]{singh_etal_16, bromberg_etal_2016, kadowaki_etal_2021, Medina-Torrejon_2021, Medina-Torrejon_2023},
and even tearing mode \citep[e.g.][]{Huang2016, Beg2022, Vicentin2024} 
\footnote{In the case of the tearing instability, however, the required level of resistivity is likely reached only once the system has already become fully turbulent \citep[][]{Vicentin2026}.}.
Unlike in PIC scales, the reconnection velocity in MHD turbulent flows is independent of resistivity. 
Instead, it depends on the scale and velocity of turbulence at the injection \footnote{The meandering of the magnetic field lines in the turbulent flow facilitates numerous simultaneous events
of reconnection and the enlargement of the outflow regions, expelling the reconnected flux very efficiently.
These two effects contribute to a reconnection rate that is a significant fraction of the Alfvén speed irrespective of the  magnetic resistivity (i.e., independent of the Lundquist number) and solely reliant on the parameters of the turbulence).}. Accelerated particles in this case can achieve energies up to approximately $10^{9}$ times the particle rest mass energy ($mc^2$), e.g. in relativistic jets of blazars with background magnetic fields $\sim 10$ G  
\citep[][]{Medina-Torrejon_2021}. The 
 dominant electric field responsible for particle acceleration is non-resistive,  given by the ($\mathbf{v} \times \mathbf{B}$) term of the turbulent magnetic fluctuations of the background flow coming into the reconnection layers.


The MHD (and several PIC) simulations along with theoretical framework have demonstrated  that particles are mainly accelerated at reconnection sites through a Fermi mechanism in  the ideal electric fields
\citep[][]{dalpino_lazarian_2005, drake_etal_2006, kowal_etal_2012, guo_etal_2019}.
In these sites, particles undergo multiple crossings in the converging magnetic fluxes of opposite polarity, gaining energy from interactions with background magnetic irregularities. The 3D simulations demonstrate the formation of these reconnecting layers throughout the turbulence's inertial range (i.e., from the small dissipation scale to the injection scale of the turbulent eddies), allowing for particle acceleration up to large scales and very high energies as remarked above \citep[][]{kowal_etal_2012, dalpino_kowal_15, delvalle_etal_16, Medina-Torrejon_2021, kadowaki_etal_2021, Medina-Torrejon_2023}.
The inherent 3D nature of turbulent reconnection and the resulting particle acceleration in 3D reconnecting flux tubes render the process more efficient compared to acceleration within 2D  shrinking plasmoids and X-points typically induced by tearing mode instability in PIC \citep[e.g.][]{hoshinoLyu2012, drake_etal_2006, sironi_spitkovsky_2014} 
 and laminar resistive magnetohydrodynamic (MHD) simulations \citep[e.g.][]{kowal_etal_2011, puzzoni_etal_2022}.

In the Fermi regime particles are  accelerated in time up to a threshold very high energy which is attained when the particle's Larmor radius reaches the injection size of the turbulence (which also determines the thickness of the largest reconnection layers).  Beyond the threshold, particles experience additional acceleration, 
albeit at a reduced rate, due to drifting in large-scale non-reconnecting fields. The energy spectrum of accelerated particles exhibits a high-energy tail with a power-law index of approximately -1 to -2 
which is influenced by both the Fermi and drift mechanisms 
\citep[][]{kowal_etal_2012, lazarian12, dalpino_kowal_15, delvalle_etal_16, Medina-Torrejon_2021, Medina-Torrejon_2023}.
\footnote{Interestingly, a recent study by \cite{Zhang_Xu2023} 
have replicated the simulation and analysis of test particle acceleration in a current sheet with self-driven turbulent reconnection and obtained results very much  similar to those reported by \cite{kowal_etal_2012, delvalle_etal_16}. 
However, they did not address the significant similarities, which successfully confirmed those earlier findings on reconnection acceleration as well.}

In contrast, 
recent 3D PIC simulations propose that drift acceleration predominantly shapes the particle spectrum over Fermi acceleration \citep[][]{sironi2022, zhang2021, Zhang2023},
although consensus on this matter regarding PIC simulations remains elusive \citep[e.g.][]{guo_etal_2019, Li_Guo2021, Guo2023}. 
Moreover, considering that drift acceleration, which is strongly dependent on the particle's energy, is much less efficient at large energies than Fermi acceleration in reaching the observed very and ultra-high energies \citep[][]{delvalle_etal_16}, it seems unlikely that this process could emerge as the only one dominant at large astrophysical scales. Hence, it is crucial to exercise caution when extrapolating results from kinetic PIC scales to larger scales of real systems.  
Still, both PIC and MHD regimes complement each other. PIC simulations have successfully investigated particle acceleration from energies below their rest mass to several hundred times this value, addressing the injection problem 
mostly for electron-positron pair plasmas \citep[e.g.][]{sironi2022, Guo2023}.
Meanwhile, MHD simulations have explored particle acceleration of protons up to the highest observed values at the macroscopic injection scales of turbulence, clearly probing the  threshold (or saturation) regime described above 
\citep[][]{kowal_etal_2012,  dalpino_kowal_15, delvalle_etal_16, Medina-Torrejon_2021, Medina-Torrejon_2023}.

Another crucial quantity for characterizing reconnection acceleration, in addition to the particle power spectrum, is the acceleration time. According to the Fermi mechanism, for a constant reconnection rate, there is an exponential energy growth over time, regardless of the energy \citep[e.g.][]{dalpino_lazarian_2005, dalpino_kowal_15}. In other words, the acceleration time is expected to be independent of particle energy, provided that the reconnection velocity, $v_{\rm rec}$, remains constant, since $\Delta E / E \propto v_{\rm rec}/c$. 
3D MHD simulations of turbulence-induced reconnection in relativistic and non-reativistic flows have shown nearly exponential growth over time, with a very weak dependence on energy, implying acceleration times approximately of $t_{acc} \propto E_p^{0.2}$ in current sheets with nearly relativistic reconnection velocities 
\citep[][]{delvalle_etal_16} and $t_{acc} \propto E_p^{0.1}$ in relativistic jets \citep[][]{Medina-Torrejon_2021, Medina-Torrejon_2023}, confirming the predictions for the Fermi regime. The slight departure from zero-dependence is attributed to the fact that in such flows, the reconnection velocity is not constant (with average values varying between $\sim 0.01 - 0.05 v_A$, where $v_A$ is the Alfvén speed). The 3D MHD simulations have also revealed a strong dependence of the particle acceleration time on the reconnection velocity \citep[][]{delvalle_etal_16},
 which is also  consistent with the theory \citep[][]{dalpino_lazarian_2005}.
As emphasized earlier, the Fermi regime continues until the particles reach a threshold energy at which their Larmor radius equals the thickness of the largest reconnection layers. Beyond  this energy, these simulations show that particles transition to slower acceleration, drifting in the gradients of non-reconnecting magnetic fields.

In a recent study, \citet[][]{xu_lazarian_2023} 
\citepalias[hereafter][]{xu_lazarian_2023} revisited the earlier work by \cite{dalpino_lazarian_2005}, who initially proposed the Fermi process as the primary mechanism for reconnection acceleration \citep[see also the review in][]{dalpino_kowal_15}. \citetalias[][]{xu_lazarian_2023} assessed the particle acceleration time in the Fermi regime within a turbulence-induced magnetic reconnection layer, delineating three distinct conditions depending on the thickness of the reconnection layer, the reconnection velocity and the angle between the reconnection and the guide fields.

In this work, we will investigate the theoretical predictions for particle acceleration time both in the Fermi 
and drift regimes, through a detailed statistical analysis of both the reconnection sites and the accelerated particles obtained from 3D MHD-PIC simulations of turbulent relativistic jets.

In \citet[][]{Medina-Torrejon_2021} \citepalias[hereafter][]{Medina-Torrejon_2021}, and \citet{kadowaki_etal_2021},
 it was investigated particle acceleration within 3D relativistic magnetically dominated jets with  magnetization parameter  $\sigma 
 \sim 1$, 
subject to  current-driven-kink instability (CDKI). 
This induces turbulence and rapid magnetic reconnection within the jet flow, leading to the disruption of the initial helical magnetic field configuration and the formation of numerous sites of fast reconnection. 
Test protons introduced into  nearly stationary turbulent snapshots of the jet undergo exponential  acceleration over time, primarily along the local magnetic field lines, reaching the aforementioned very and ultra-high energies even before exiting the reconnection layers. Accelerated particles show a clear association with regions of fast reconnection and high current density. In the subsequent work by \citet[][]{Medina-Torrejon_2023} \citepalias[hereafter][]{Medina-Torrejon_2023},   3D MHD-PIC simulations of the same system were performed (considering also $\sigma  \sim 10$), 
in order to scrutinize the early stages of particle acceleration simultaneously with the growth of turbulence by CDKI in the jet. Instead of injecting particles in the nearly steady state snapshots of the simulated turbulent jet, they were injected from the beginning into the system, allowing them to evolve with the jet. These simulations corroborated the prior findings, illustrating the significant potential of magnetic reconnection driven by turbulence to propel relativistic particles to exceedingly high energies within magnetically dominant flows. 

In this work we will employ the \citetalias[][]{Medina-Torrejon_2021} and \citetalias[][]{Medina-Torrejon_2023} simulations of the $\sigma \sim 1$ jet to probe the reconnection acceleration time predictions outlined by \citetalias[][]{xu_lazarian_2023} for the Fermi  regime, and then compare them with the acceleration times derived independently from the accelerated particles  in \citetalias[][]{Medina-Torrejon_2021} and \citetalias[][]{Medina-Torrejon_2023}. We will also probe the theoretical predictions for the subsequent drift regime which occurs when particles reach the threshold energy of the Fermi regime.


The plan of the paper is as follows. In Section \ref{tacc_theory}, we summarize the theoretical predictions for the reconnection acceleration time, in section \ref{sec:style} we describe the numerical method and  setup of the simulations, in Section \ref{sec:results}, we present the results and in Section \ref{sec:discussion} we discuss the results and draw  our conclusions.

\section{Theoretical Predictions for  reconnection Acceleration Time}
\label{tacc_theory}

The acceleration time resulting from magnetic reconnection is ultimately constrained by the size of the acceleration region divided by the reconnection velocity ($v_{rec}$). In the case of turbulence-induced reconnection, $v_{rec}$ is a substantial fraction of the Alfvén speed \citep[][]{lazarian_vishiniac_99, kowal_etal_09, delvalle_etal_16,  kadowaki_etal_2021, Vicentin2024}.

As noted in Section \ref{sec:intro}, during the Fermi regime of particle acceleration through reconnection, it is expected that particles will experience exponential growth in energy over time \citep[e.g.][]{dalpino_lazarian_2005}. MHD simulations with test particles support this expectation, showing a reconnection acceleration time with a very weak dependence on particle energy \citep[e.g.][]{delvalle_etal_16, 2017PhRvL.118h5101L},  and \citepalias[][]{Medina-Torrejon_2021, Medina-Torrejon_2023}.

In  recent study, \citetalias{xu_lazarian_2023} revisited the earlier work by \citet{dalpino_lazarian_2005} and derived the following conditions for the acceleration time in the Fermi regime within a turbulence-induced magnetic reconnection layer:
\begin{eqnarray}
\textup{Condition I}\, (\textup{small}\,\, \theta, \textup{small}\,\, v_{in}):\,\,\, &t_{acc}& \sim \frac{3 \Delta}{\sin \theta v_{in}} , \label{tacc1} \\ 
\textup{Condition II}\,(\textup{large}\,\, \theta, \textup{small}\,\, v_{in}):\,\,  &t_{acc}& \sim \frac{3 \Delta}{2 v_{in}} ,\label{tacc2} \\ 
\textup{Condition III}\,(\textup{large}\,\, \theta, \textup{large}\,\, v_{in}):\,\,  &t_{acc}& \sim \frac{4 \Delta}{c d_{ur}} , \label{tacc3}
\end{eqnarray}
where $\Delta$  is the thickness of the reconnection layer, $v_{in}$ is the inflow or  reconnection speed in units of the light speed $c$,  
$\theta$ is the angle between the inflow
magnetic field to the reconnection layer and the guide field, and:
\begin{equation}
d_{ur} \approx \frac{2\beta_{in}(3\beta^2_{in}+3\beta_{in}+1)}{3(\beta_{in}+ 0.5)(1-\beta^2_{in})},
\end{equation}
being  $\beta_{in} = v_{in}/c$ and $c$  the light speed.

In the subsequent drift regime, that occurs  after particles attain the threshold energy ($E_{p,th}$) and exit the reconnection site, or in other words, when their  Larmor radius $r_L \geq \Delta$ \citep[][]{kowal_etal_2012, lazarian12},  the energy growth with time becomes  strongly dependent on the energy.  The acceleration time in this regime is approximately described by  
\citep[][]{dalpino_kowal_15, delvalle_etal_16, zhang2021, Zhang2023}:
\begin{eqnarray}
t_{acc, drift} \simeq \frac{E_p}{q B v_{rec}},  \,\,\,E_p\geq E_{p,th}
\label{adrift}
\end{eqnarray}
where $E_p$ is the particle energy, B the magnetic field, and $q$  the charge particle. All of those parameters are given in cgs units.

3D  MHD numerical simulations with test particles have confirmed that the extended acceleration time observed in the drift regime is attained only for $E_p > E_{th}$  \citep[][]{kowal_etal_2012, delvalle_etal_16}, \citepalias[][]{Medina-Torrejon_2021}. This regime has been also detected recently in 3D PIC simulations \citep[][]{Li_Guo2021,zhang2021, Zhang2023}.

\section{Numerical method and setup} 
\label{sec:style}

\begin{table*}
\centering
\caption{Parameters of the test particle models.}
\label{tablepartic}
\begin{tabular}{c cc cc}
\firsthline
\hline
Model & Jet time 
& L [pc]      & Particles evolution time \\
\hline
RMHD-PIC    & 0 - 60 L/c 
&  $5.2 \times 10^{-7}$ 	& $\sim 1$ hr.  \\ 

RMHD-GACCEL    & 45 L/c	
& $3.5 \times 10^{-5}$	& $\sim 10^4$ hr.     \\ 
\lasthline
\end{tabular}
\end{table*}
For the present analysis we employ the same relativistic  simulation described in  \citetalias[][]{Medina-Torrejon_2023} for a jet with magnetization $\sigma\sim$ 1. That simulation utilized the relativistic  MHD-PIC version of the  \texttt{PLUTO} code \citep{mignone_etal_2018}, with a resolution $256^3$ in a box of $10L \times 10L \times 6L$ where $L$ is the length scale unit. An initial helical magnetic field configuration with maximum  intensity $B_0= 0.7$ code unit (c.u.) at the central axis is assumed. Its profile is given in \citetalias{Medina-Torrejon_2021}.
The  initial density, pressure, and magnetization parameter  at the jet axis (at the plasma frame) are $\rho = 0.8$ c.u., $p = 0.02$ c.u., and   $\sigma_0 = B_0^2/\gamma^2 \rho h \sim 0.6$, respectively, where $\gamma$ is the Lorentz factor and $h$ is the specific enthalpy (with $\gamma \sim 1$ and $h \sim 1$ at the axis). The code unit   for  density  is $\rho_0 = $1, for velocity is the light speed  $c$, for time is $L/c$,  for  magnetic field is  $\sqrt{4 \pi \rho_0 c^2}$,  and for  pressure is $\rho_0 c^2$. An initial perturbation is applied to allow the growth of the CDKI and turbulence \citepalias[see][for details]{Medina-Torrejon_2021}. The simulation is run up to t= 60 L/c. 

50,000 particles were injected in this domain at t=0, with a charge to mass ratio given by $e/mc =$ 20,000 (which relates to the physical ratio as $e/mc = (e/mc)_{cgs} L_{cgs} \sqrt{\rho_{cgs}}$).  This corresponds to a physical length unit for the system $L  \sim 5.2 \times 10^{-7}$ pc, for  a jet density unit $\rho_{cgs} \sim 1.67 \times 10^{-24}g/cm^3$ (or 1 particle/$cm^{3}$).
The corresponding initial maximum magnetic field intensity in physical units is $0.1$ G \citepalias[][]{Medina-Torrejon_2021, Medina-Torrejon_2023}.
Particles were initially distributed uniformly across the domain with initial kinetic energies ranging from $(\gamma_p - 1) \sim$ 1 to  200, where $\gamma_p$ is the particle Lorentz factor, with velocities randomly assigned following a Gaussian distribution \citepalias[see][for more details]{Medina-Torrejon_2023}. 
In  Table \ref{tablepartic} this test-particle model is referred as RMHD-PIC. The table gives the dynamical  time evolution of the jet, its physical size unit, and the corresponding  evolution time of the particles in this system.  

In addition, in our  analysis below we also consider a post-processing test particle simulation also  taken from \citetalias[][]{Medina-Torrejon_2023} \citepalias[see also][]{Medina-Torrejon_2021}. In this case,  one thousand test-particles were injected in a fully developed turbulent snapshot of the same jet system 
(at t=45 L/c), 
and allowed to accelerate for much longer time (see Figure 5, bottom, in \citetalias{Medina-Torrejon_2023})\footnote{We note that in the case of post-processing test-particle simulations, the number of injected particles can be much smaller than in the RMHD-PIC model. This has been thoroughly tested in \citetalias{Medina-Torrejon_2021} who found no major improvement in the statistical analyses by increasing the injected particles' number by factors of 10 to 20.}. As in \citetalias[][]{Medina-Torrejon_2021} and \citetalias{Medina-Torrejon_2023}, the adopted physical length scale of the system in this model was $L \sim 3.5 \times 10^{-5}$ pc, and test particle acceleration was performed with the  \texttt{GACCEL} code \citep[][]{kowal_etal_2012}\footnote{https://gitlab.com/gkowal/gaccel}.  This model is referred  as RMHD-GACCEL in Table \ref{tablepartic}, which also gives the single dynamical jet snapshot considered in this case, the corresponding physical unit length scale and the total evolution  time of the particles. 
Furhter details on the initial setup of this model can be found in \citetalias{Medina-Torrejon_2023}.

\subsection{Calculation of the reconnection parameters in the current sheets} \label{sec:params} 

In order to probe the regimes of particle acceleration in a reconnection site as described by eqs. \ref{tacc1} to \ref{tacc3}, we use the magnetic reconnection search algorithm developed in \citet[][]{kadowaki_etal_2021} \citep[see also][]{Kadowaki_2018}, 
which allows to identify all the reconnection layers in the system as well as quantify their properties including the thickness $\Delta$ and the reconnection velocity. 

In eqs. \ref{tacc1} to \ref{tacc3}, we consider (see Appendix \ref{appendix:rec_rate}):
\begin{eqnarray}
\theta = \left\{\begin{matrix}
\textup{small}\,\,\, (\sin \theta < \left \langle  \sin \theta \right \rangle )  \\ 
\textup{large}\,\,\, (\sin \theta \geq  \left \langle  \sin \theta \right \rangle )
\label{array1}
\end{matrix}\right.
\end{eqnarray}

\begin{eqnarray}
v_{in} = \left\{\begin{matrix}
\textup{small}\,\,\, (v_{in} <  \left \langle v_{in}  \right \rangle)  \\ 
\textup{large}\,\,\, (v_{in} \geq  \left \langle v_{in}  \right \rangle)
\label{array}
\end{matrix}\right.
\end{eqnarray}

\noindent where $\left \langle v_{in}  \right \rangle$ $ =\left \langle v_{\rm rec}  \right \rangle$ is the average reconnection velocity in the simulated turbulent jet, which  is of the order of  $\left \langle v_{rec}  \right \rangle \sim 0.03 v_A$, where $v_A$ is the local Alfvén speed    \citepalias{Medina-Torrejon_2023}.
We observe that in turbulence-driven reconnection, a fast reconnection speed is achieved once turbulence is fully developed. At turbulence saturation, the  reconnection speed also attains its maximum average  value \citep[e.g.][]{kowal_etal_09, kadowaki_etal_2021, Vicentin2024}. The  value above has been computed from the distribution evolution of the reconnection events in our relativistic jet \citepalias[][]{Medina-Torrejon_2023}
(see Figure \ref{hist-delta-theta-figure} in the Appendix \ref{appendix:rec_rate}). Previous studies based on 3D MHD and relativistic MHD simulations of turbulence-induced reconnection have found similar average values for fast reconnection speeds \citep[e.g.][]{singh_etal_16,  Kadowaki_2018, kadowaki_etal_2021,  kowal_etal_2020, Beg2022, Wang2023, Vicentin2024}. Moreover,  previous studies of test particle acceleration indicate that reconnection  events with rates $v_{rec}\gtrsim <v_{rec}>$  are the ones effectively capable of accelerating particles by reconnection \citep[][]{kowal_etal_2012, delvalle_etal_16, Medina-Torrejon_2021, Medina-Torrejon_2023}.
\footnote{
 We note that in \cite{xu_lazarian_2023}, the condition $v_A \sim c$ is used only as a reference threshold, which is consistent with our study  where $v_A$ takes values of $v_A > 0.3\,c$. Additionally, their choice of $v_{\text{in}}$ values was illustrative, not intended as physically realistic estimates based on turbulence-driven reconnection theory and simulations. In their model, $v_{\text{in}}$ is derived from their Eq. 6, where the parameter $f$, which encapsulates the influence of turbulence, is left unspecified. As a result, the formulation can accommodate a wide range of $v_{\text{in}}/v_A$ ratios depending on the turbulence characteristics of the system.
}

To compute $\theta$ in eqs. \ref{tacc1} to \ref{tacc3}, we  consider  the angle between the local inflow magnetic field to the reconnection layer and the local background field. 

The values of  $\Delta$, $v_{in}$, $\theta$ and $\sin \theta$, of the several reconnection sites that arise once turbulence develops are obtained directly with the reconnection search algorithm at each snapshot. 
A detailed statistical analysis of the distribution of  the  parameters  $\Delta$, and $v_{in}$, 
was already presented in \citep[][]{kadowaki_etal_2021} for a similar  simulated jet background. Nevertheless, for completeness, we present in  Appendix \ref{appendix:rec_rate}  the distribution evolution of $\Delta$, $\sin\theta$, and $\theta$,  which can be directly compared to the $v_{rec}$ distribution, which is the same as depicted in Figure 4 (top)  of \citetalias{Medina-Torrejon_2023} for this model. We refer to \citep[][]{kadowaki_etal_2021} for more details. These distributions are used to compute the acceleration time  from eqs. \ref{tacc1} to \ref{tacc3} within the reconnection layers, in Section \ref{sec:results}.


\section{Results} \label{sec:results}

\begin{figure*}
 \centering
 \includegraphics[scale=0.21]{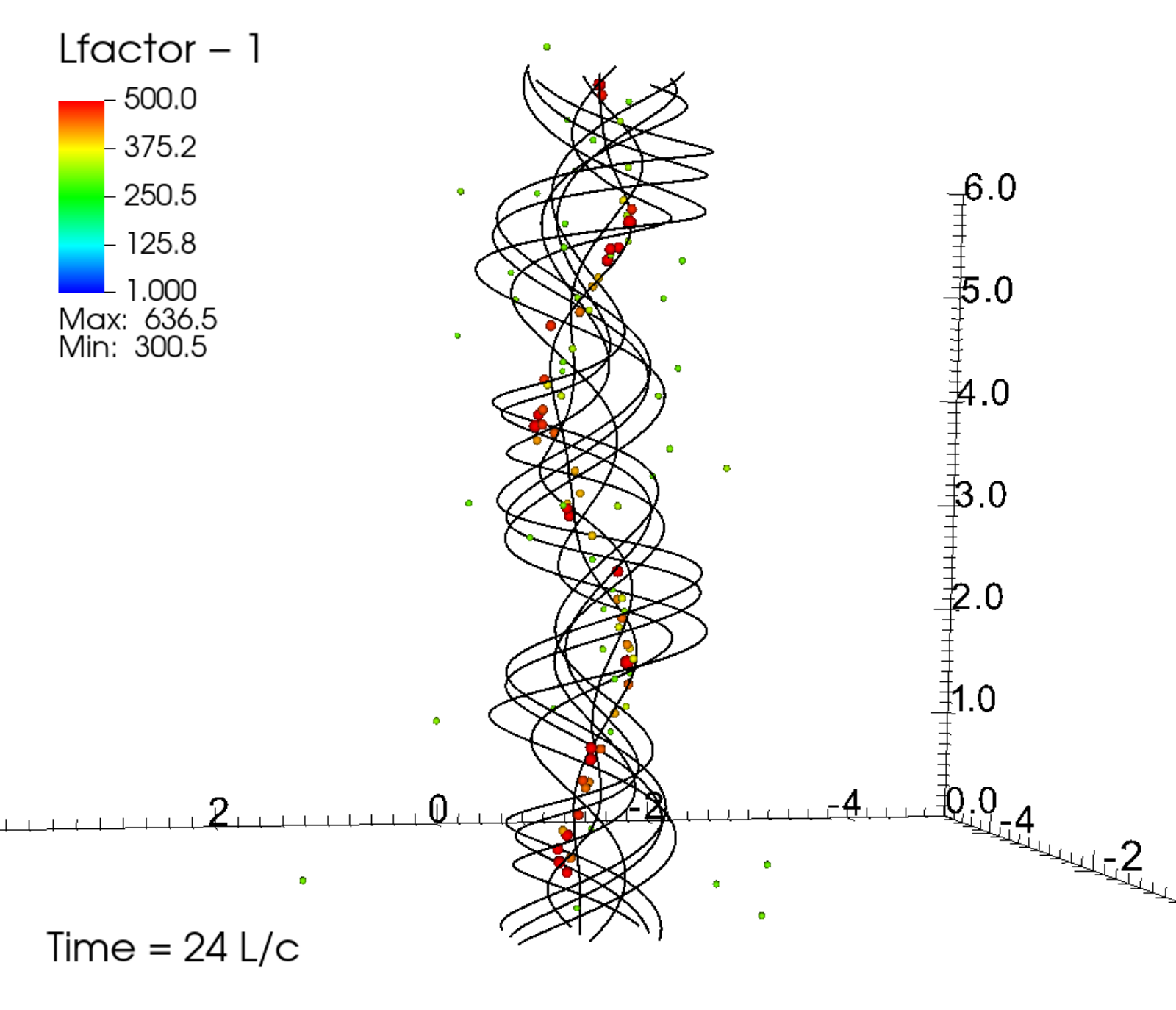}
 \includegraphics[scale=0.21]{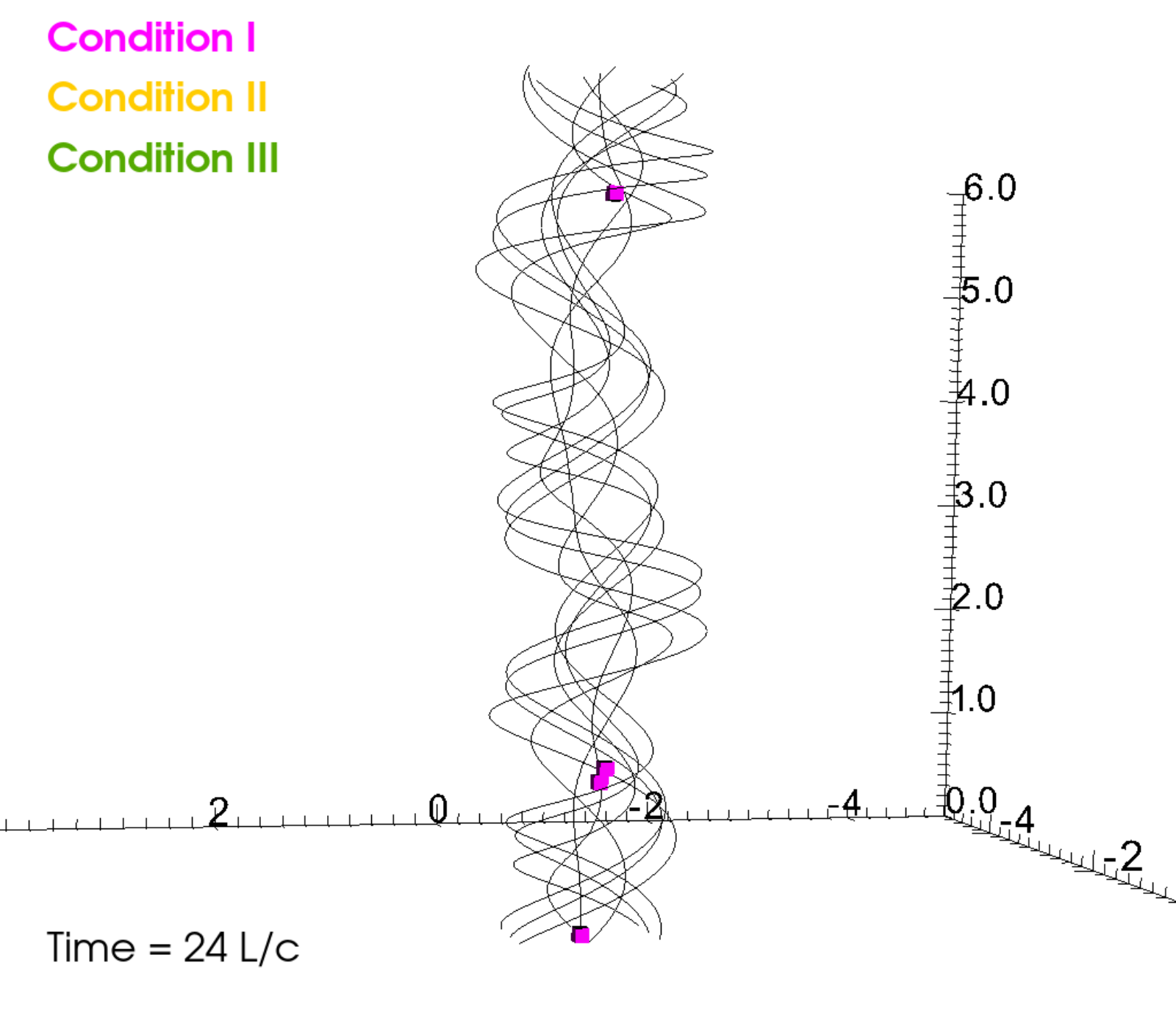}
 \includegraphics[scale=0.21]{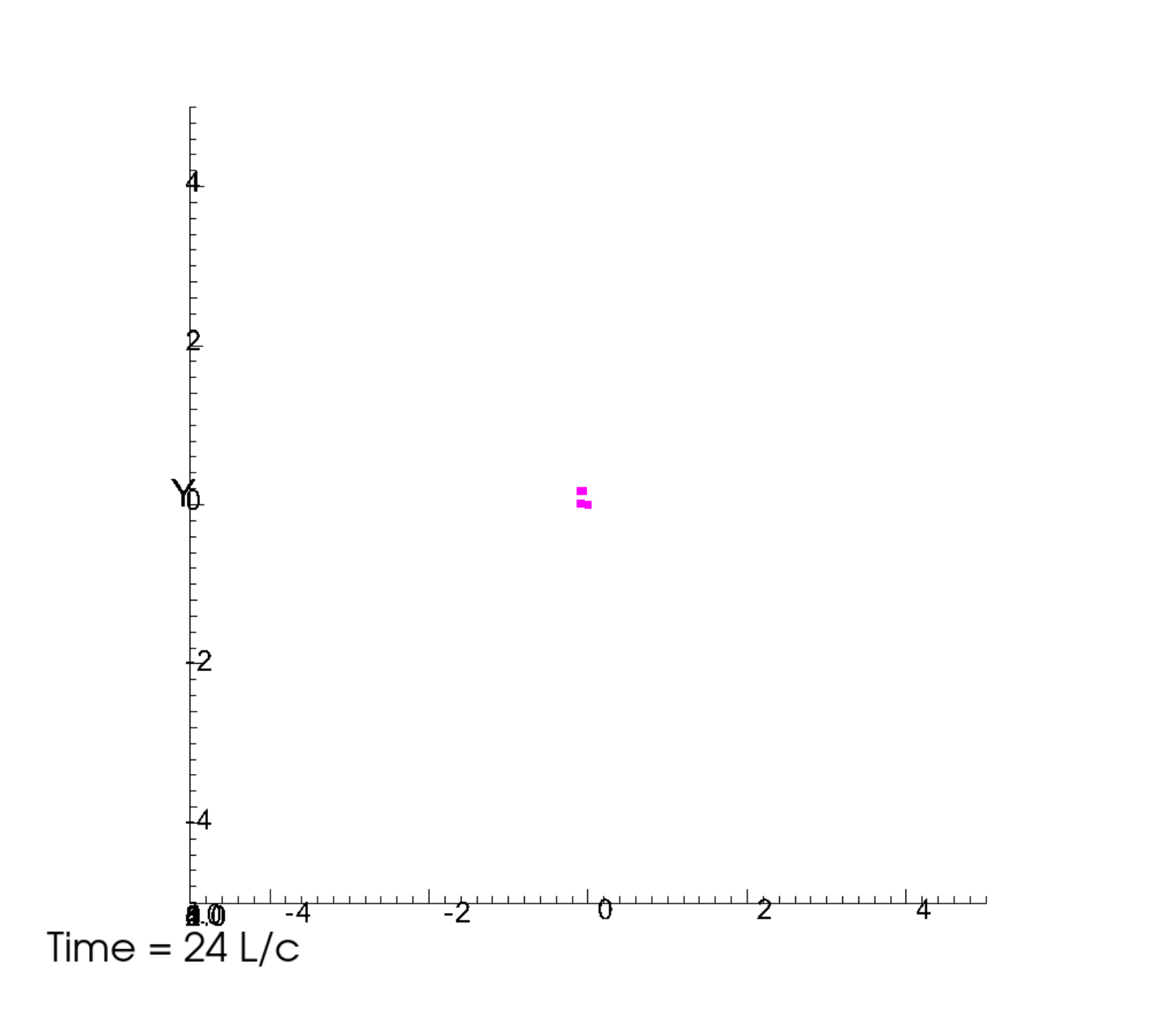}
 \includegraphics[scale=0.21]{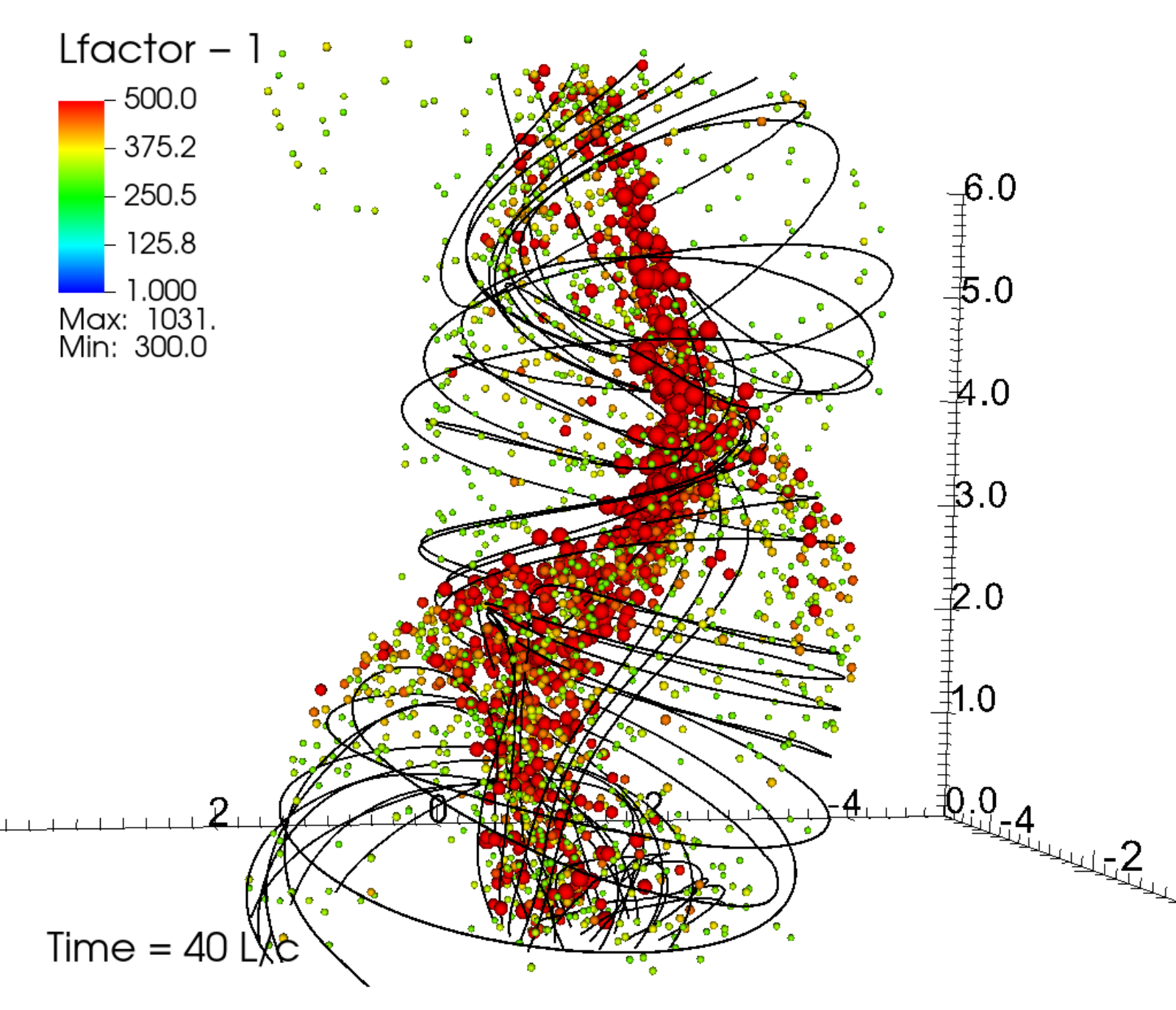}
 \includegraphics[scale=0.21]{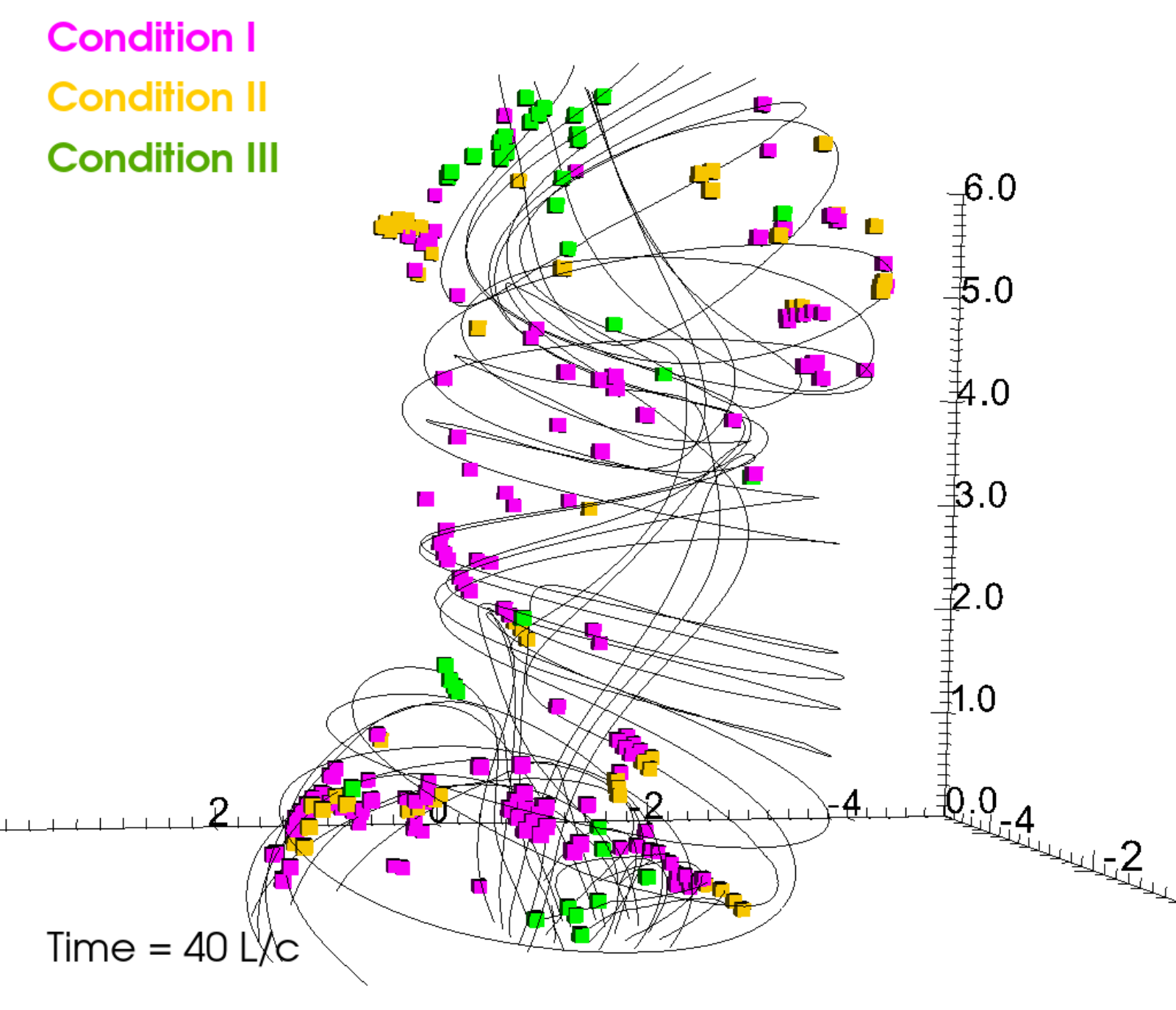}
 \includegraphics[scale=0.21]{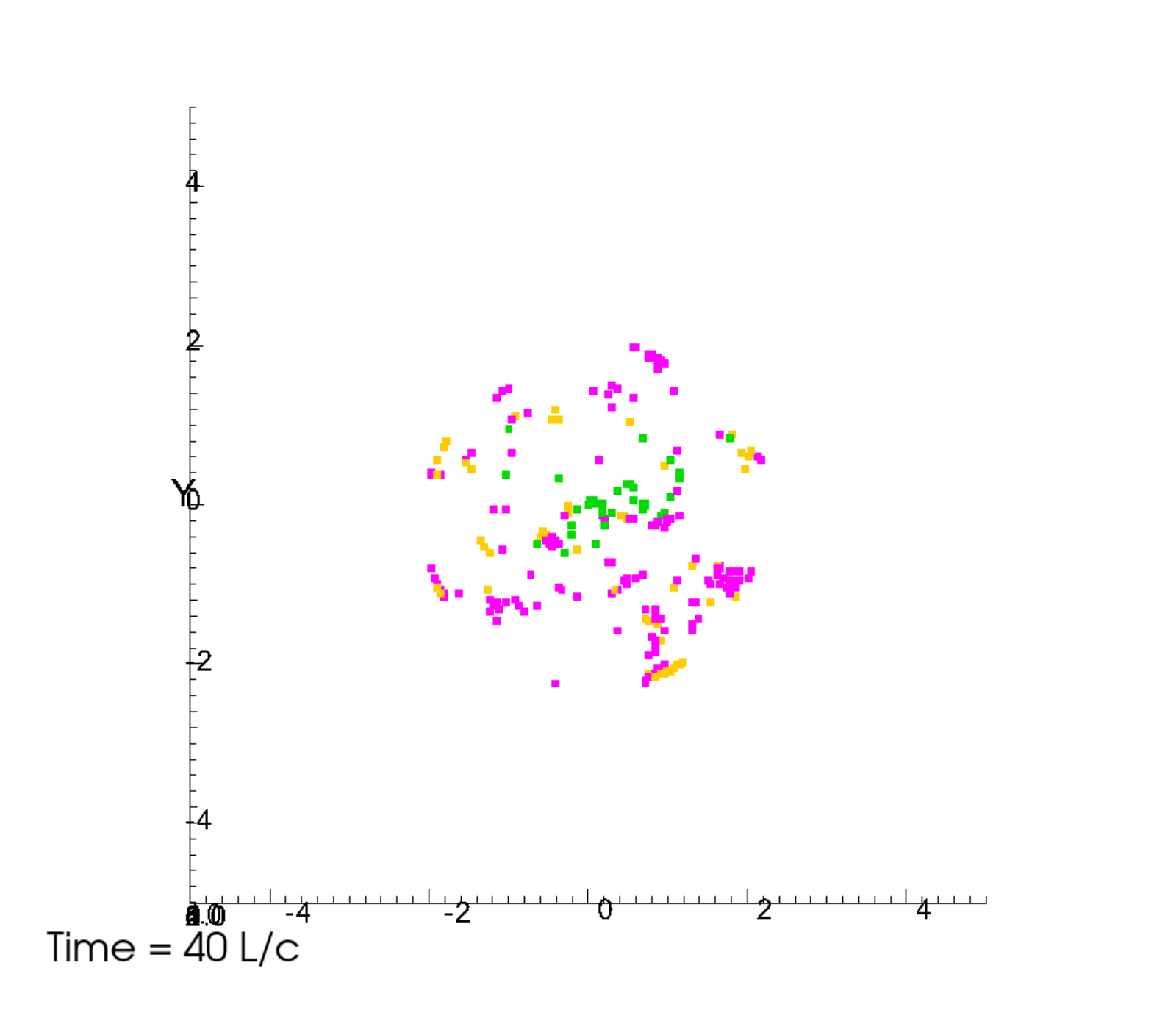}
 \includegraphics[scale=0.21]{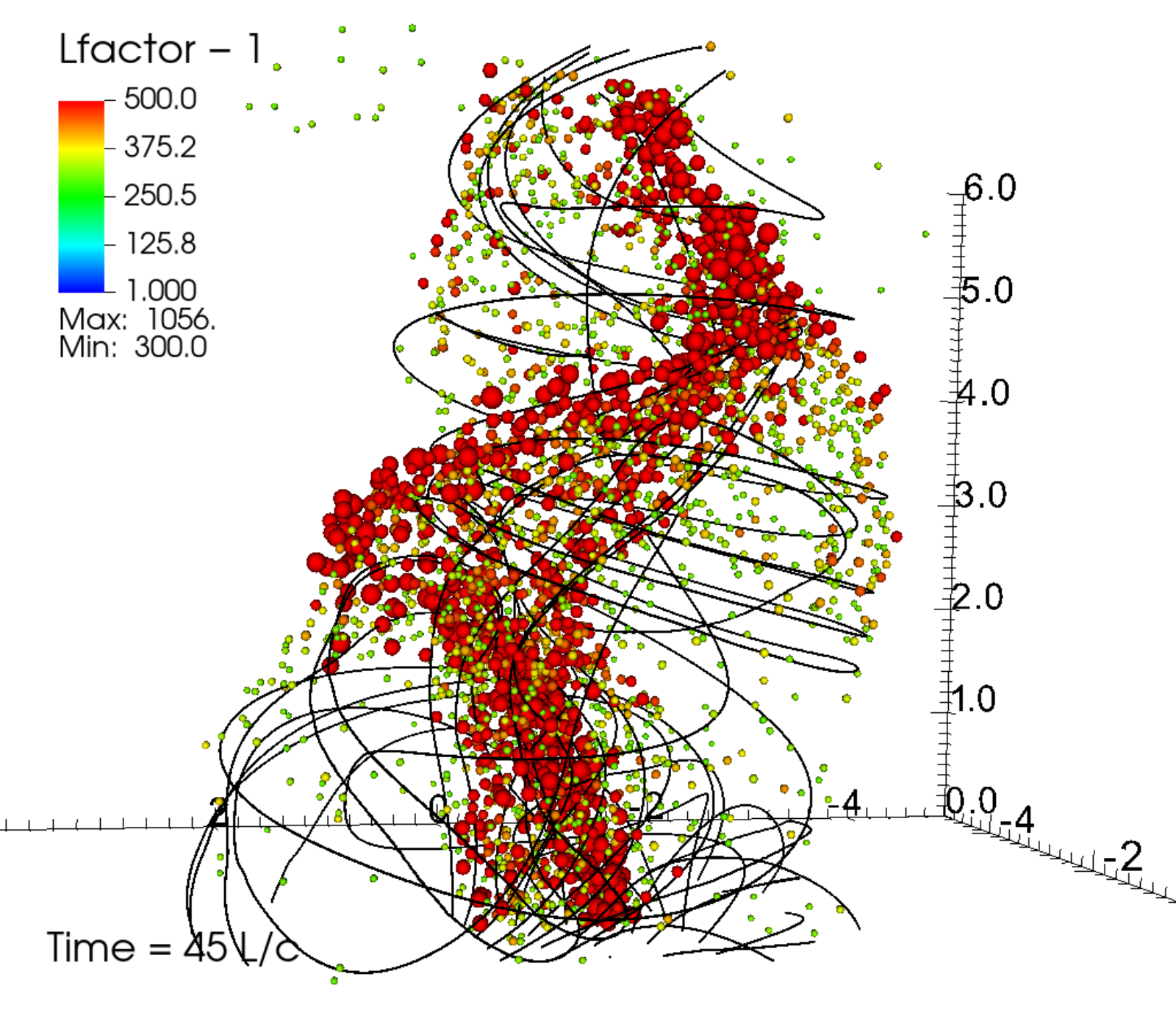}
 \includegraphics[scale=0.21]{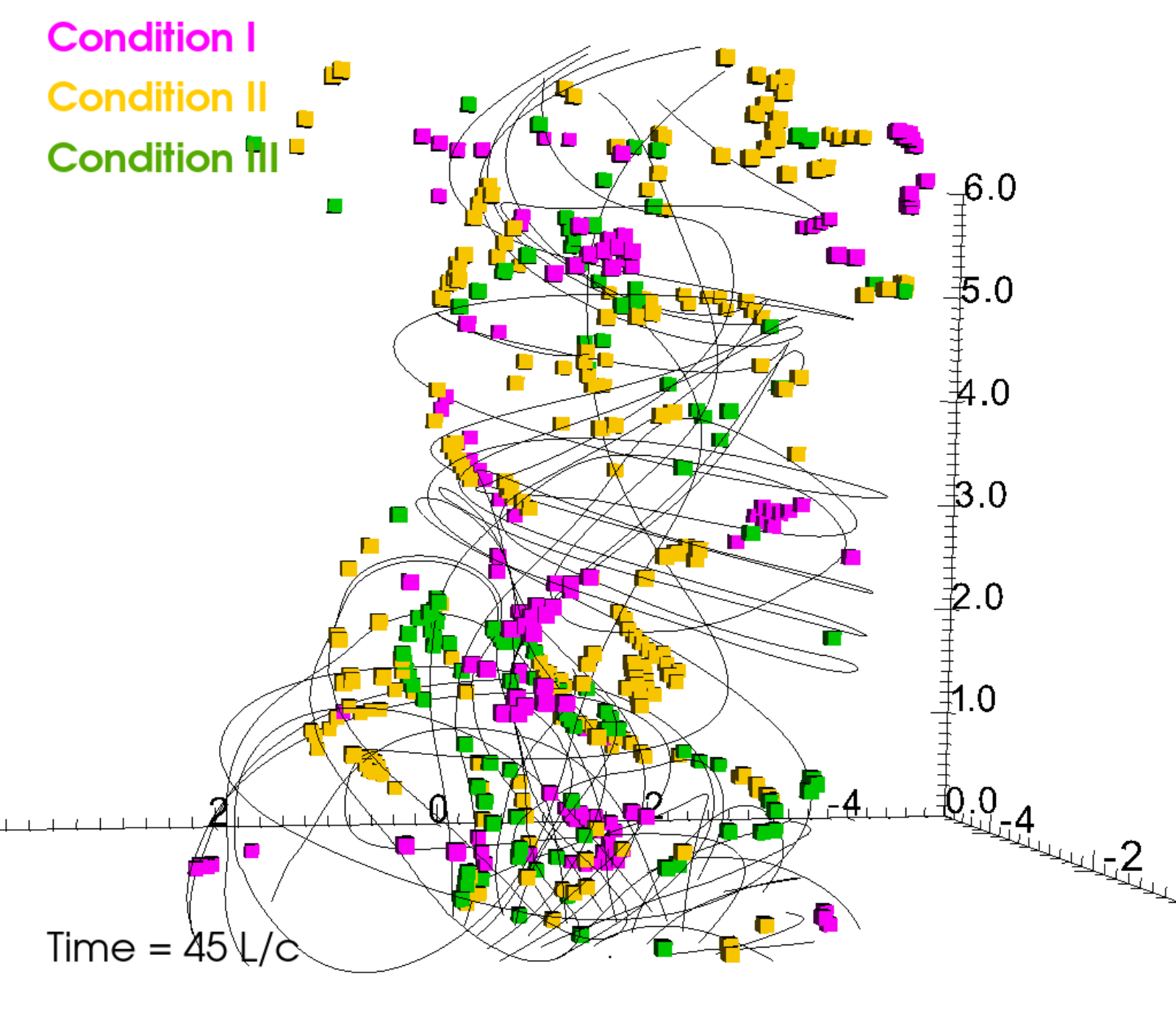}
 \includegraphics[scale=0.21]{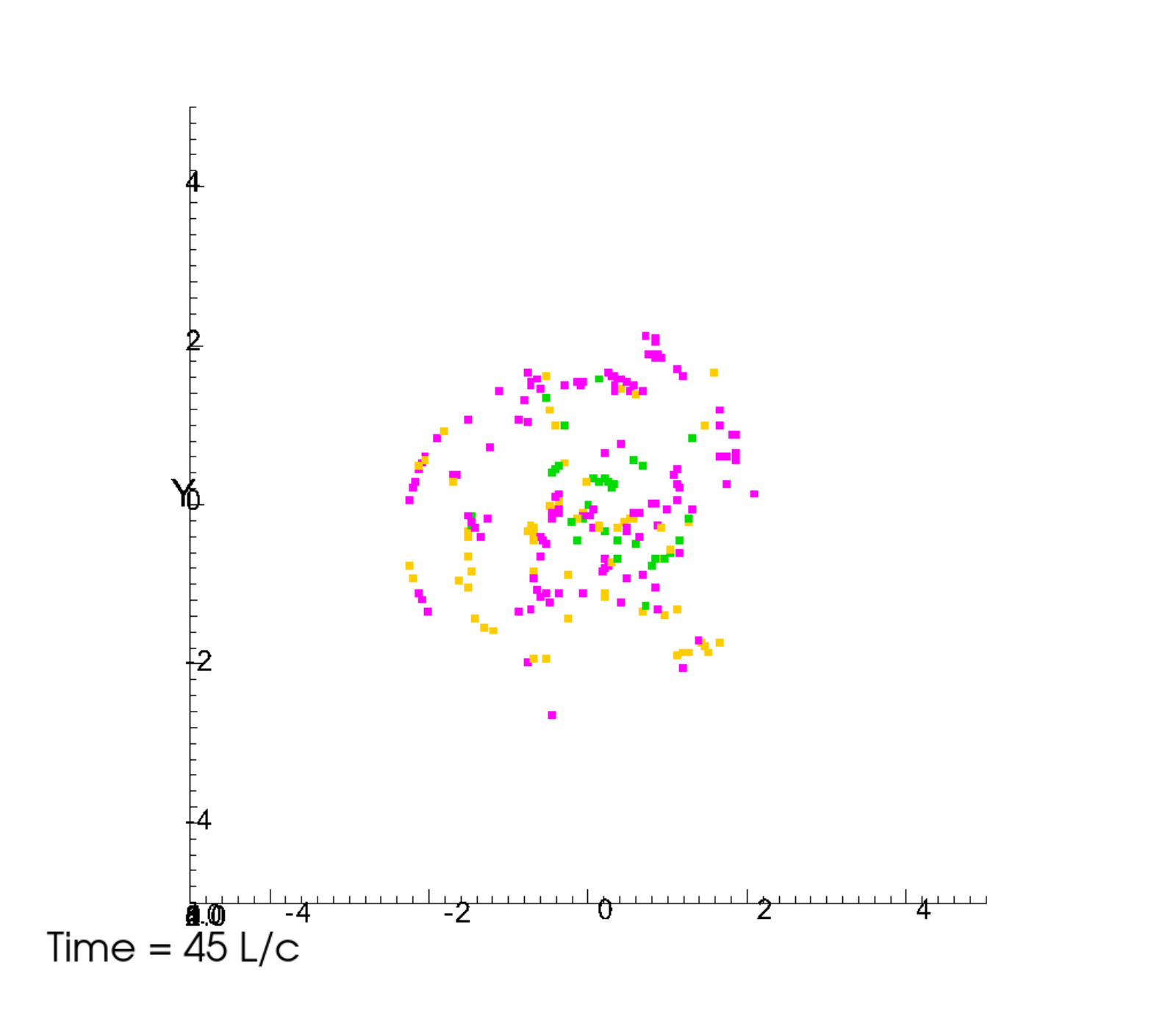}
\caption{ A 3D view of the simulated jet  at three
different snapshots   $t = 24$, $40$ and  $45$ L/c. In both  panels  the black lines represent the magnetic ﬁeld. In the left panels, 
the circles represent the particles kinetic energy normalized by the rest mass energy ($\gamma_p -1$). Only particles with energy larger than or equal to 300 are depicted.  The growing circles reflect increasing energy as well as  the change of colors (from green to red), 
which are depicted in the detail. In the middle panels the squares correspond to the central positions of all the magnetic reconnection events, the colors stand for the three different conditions as described in eqs. \ref{tacc1} to \ref{tacc3}. The rightmost panels display the same reconnection sites but in an XY (face-on) projection.
} \label{jet3D}
\end{figure*}

\begin{figure*}
    \centering
    \includegraphics[width=0.4\linewidth]{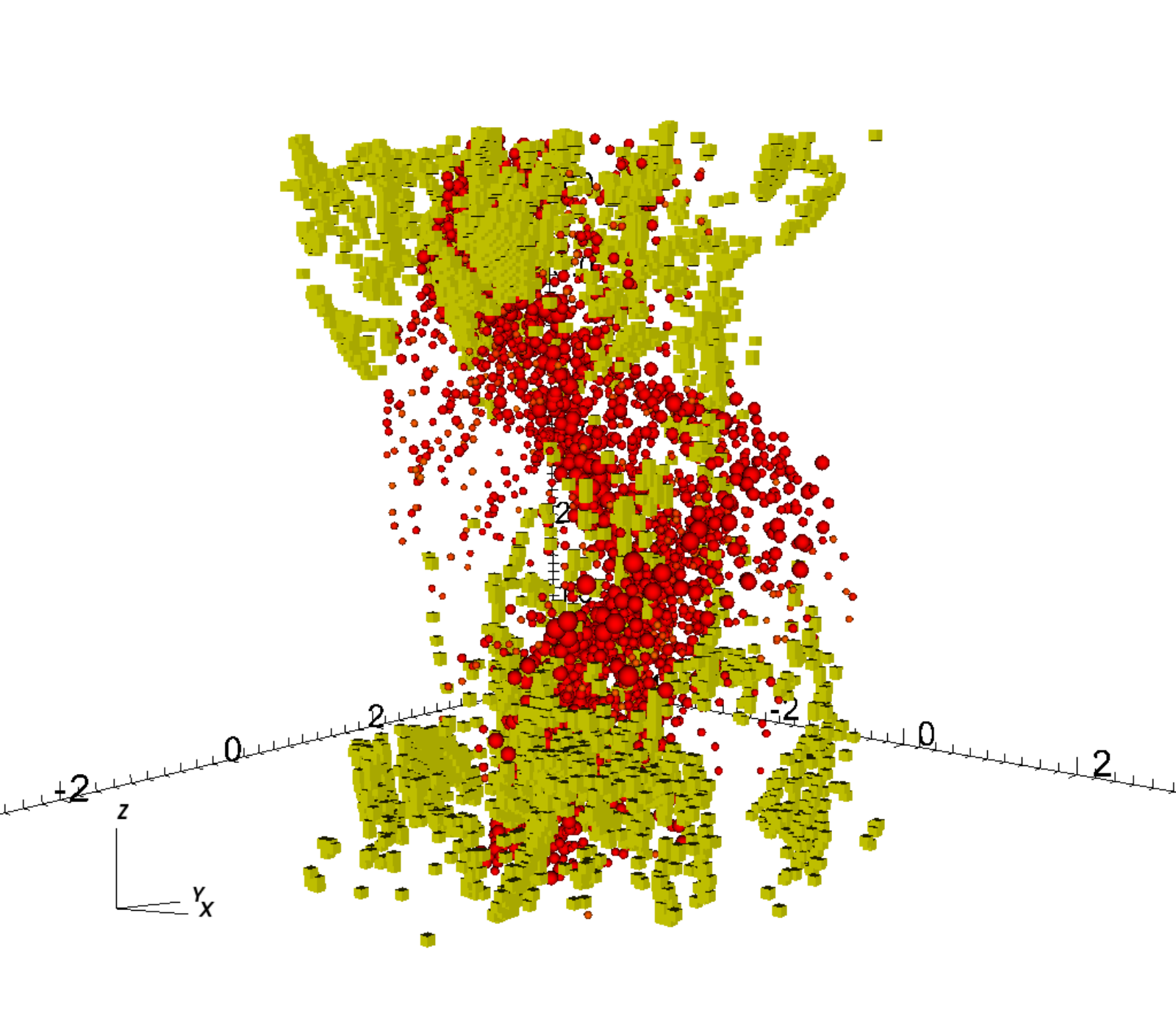}
    \includegraphics[width=0.4\linewidth]{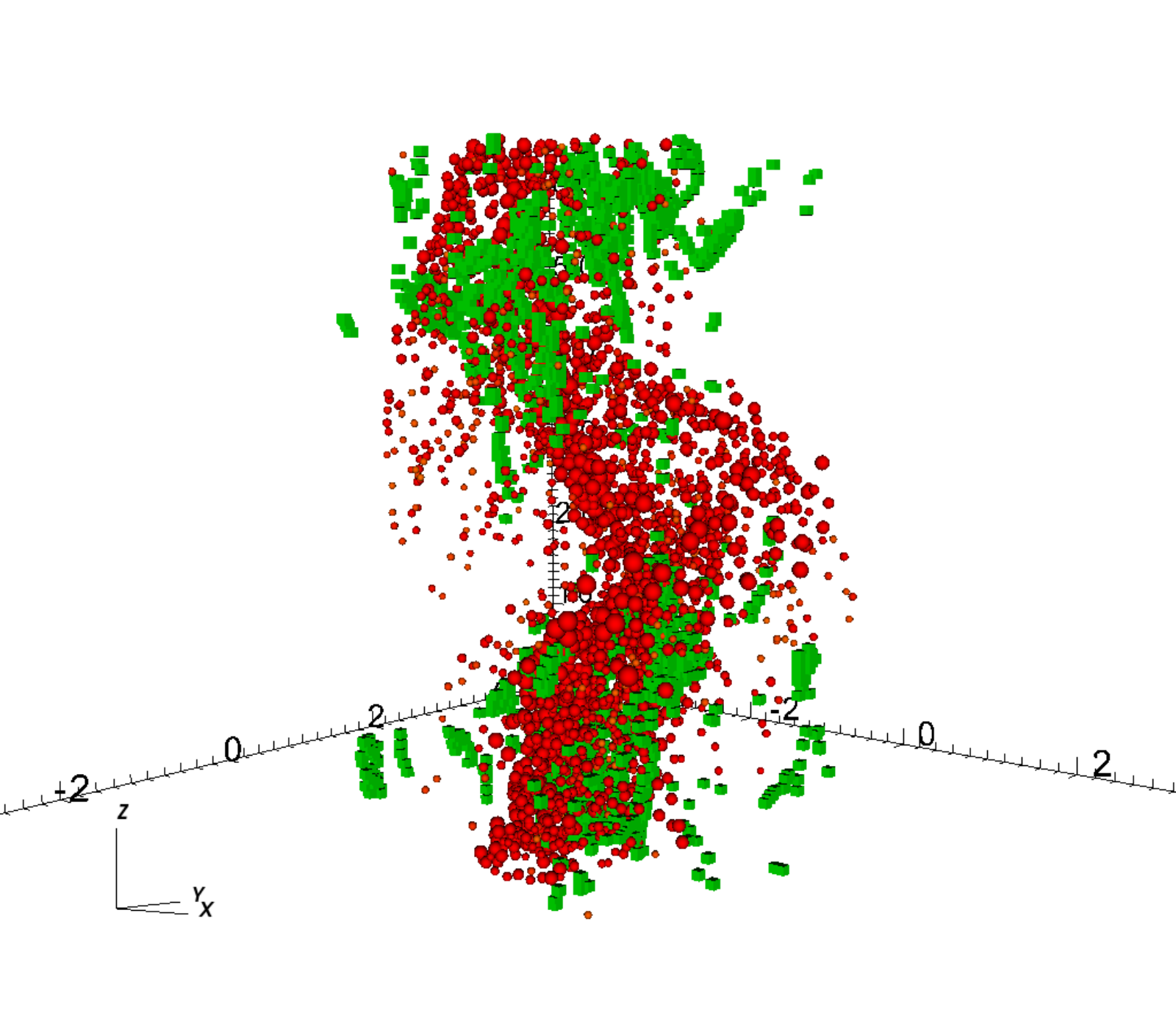}
    \caption{The left panel shows the reconnection sites accumulated from t=35 to 45 L/c for condition II, in yellow, while the right panel shows those accumulated over the same interval for condition III, in green. The red circles represent particles with energies equal or higher than 400 at t=35, 40, and 45,L/c.}
    \label{fig:placeholder35-45}
\end{figure*}

Figure \ref{jet3D}, shows 3D views of different snapshots of the jet. Each snapshot presents the particles  being accelerated in the RMHD-PIC model, superimposed on the magnetic field lines (left diagrams) and the magnetic reconnection sites for the three conditions  identified by the \cite{kadowaki_etal_2021} reconnection search algorithm (middle and right diagrams). The top  snapshot on the left shows the system when the CDKI starts to grow and the jet column is deformed in zigzags driving an initial particle acceleration dominated by curvature drift, as shown in \citetalias{Medina-Torrejon_2023} \citep[see also][]{alves_etal_2018} and  \citepalias{Medina-Torrejon_2021}. In this phase, we detect only very few incidental reconnection sites with 
slow
acceleration time values as shown  on the middle  diagram. According to the results in \citetalias{Medina-Torrejon_2023}, turbulence driven by the CDKI develops in the system only after $t=30$ L/c.
The middle and bottom snapshots, at $t= 40$ and $45$ L/c, respectively, highlight the  appearance of several particles being accelerated in the exponential regime  (see the histogram of the particles energy distribution evolution with time in the upper panel of Figure 6 in \citetalias{Medina-Torrejon_2023}). They coincide with the emergence of several reconnection sites  induced by fully developed turbulence in these snapshots.  
The rightmost panels of Figure \ref{jet3D} show the distribution of the reconnection events projected onto the XY plane (face-on view) for each snapshot. At t=40 and 45 L/c, the events with the longest acceleration times (condition I, pink) are preferentially located near the jet periphery, whereas the fastest events (condition III, green) are more concentrated toward the central region. The intermediate cases (condition II, yellow) are mostly distributed between these two extremes. This spatial distribution is consistent with that of the accelerated particles, which are concentrated mainly along the distorted jet spine and show a clearer association with the fastest reconnection events, especially those of condition III, although associations with the more numerous condition II events are also present. To illustrate this correspondence more clearly, Figure \ref{fig:placeholder35-45} shows separately the reconnection sites accumulated from t=35 to 45 L/c for condition II (left panel) and condition III (right panel), superposed on the distribution of accelerated particles with energies greater than or equal to 400 $mc^2$. The figure shows that the accelerated particles accumulate predominantly along the jet spine and are more strongly associated with the fastest reconnection events, i.e., those with the shortest acceleration times (green) (see also Figure 1 of \citetalias{Medina-Torrejon_2023}).



\begin{figure}
 \centering
   \includegraphics[scale=0.4]{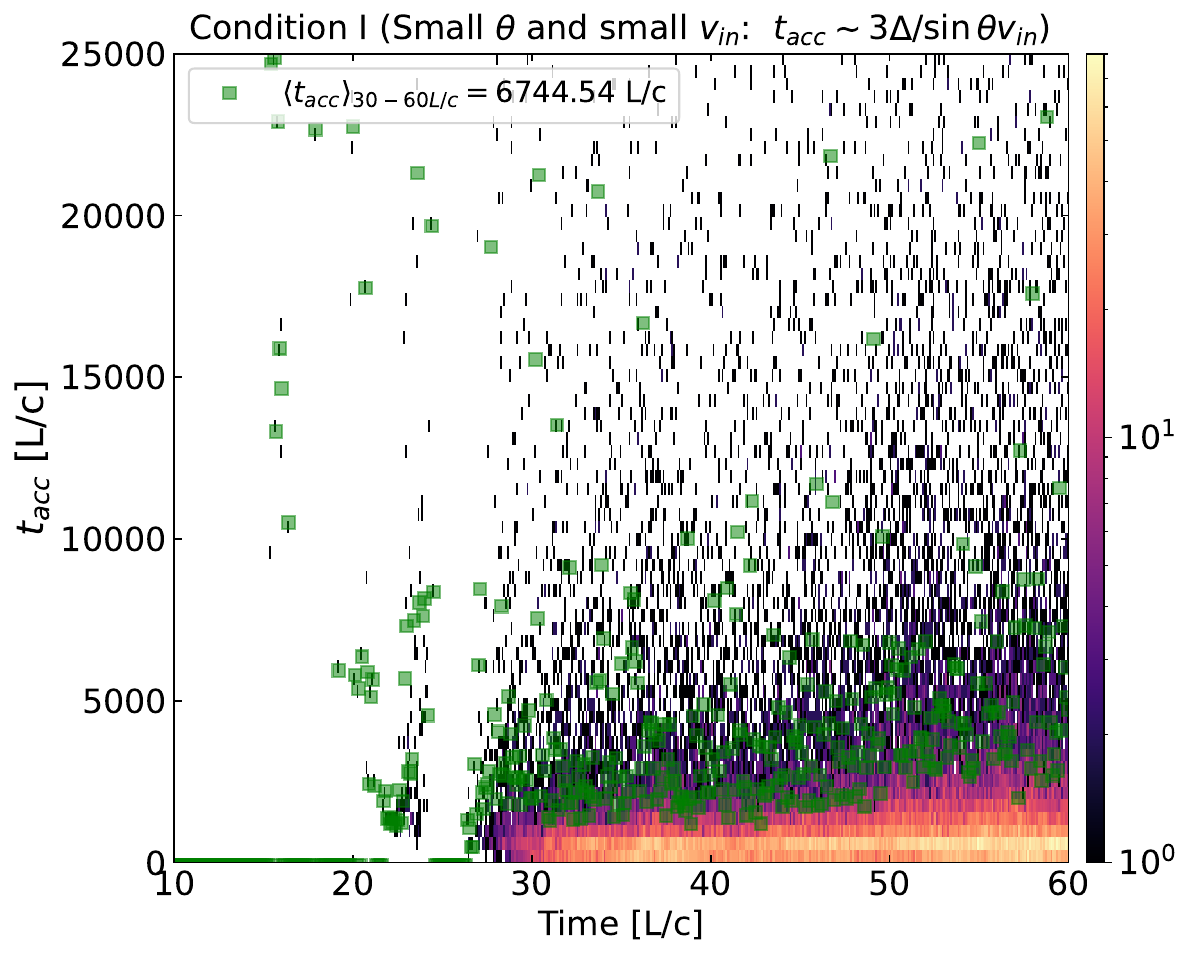}
   \includegraphics[scale=0.4]{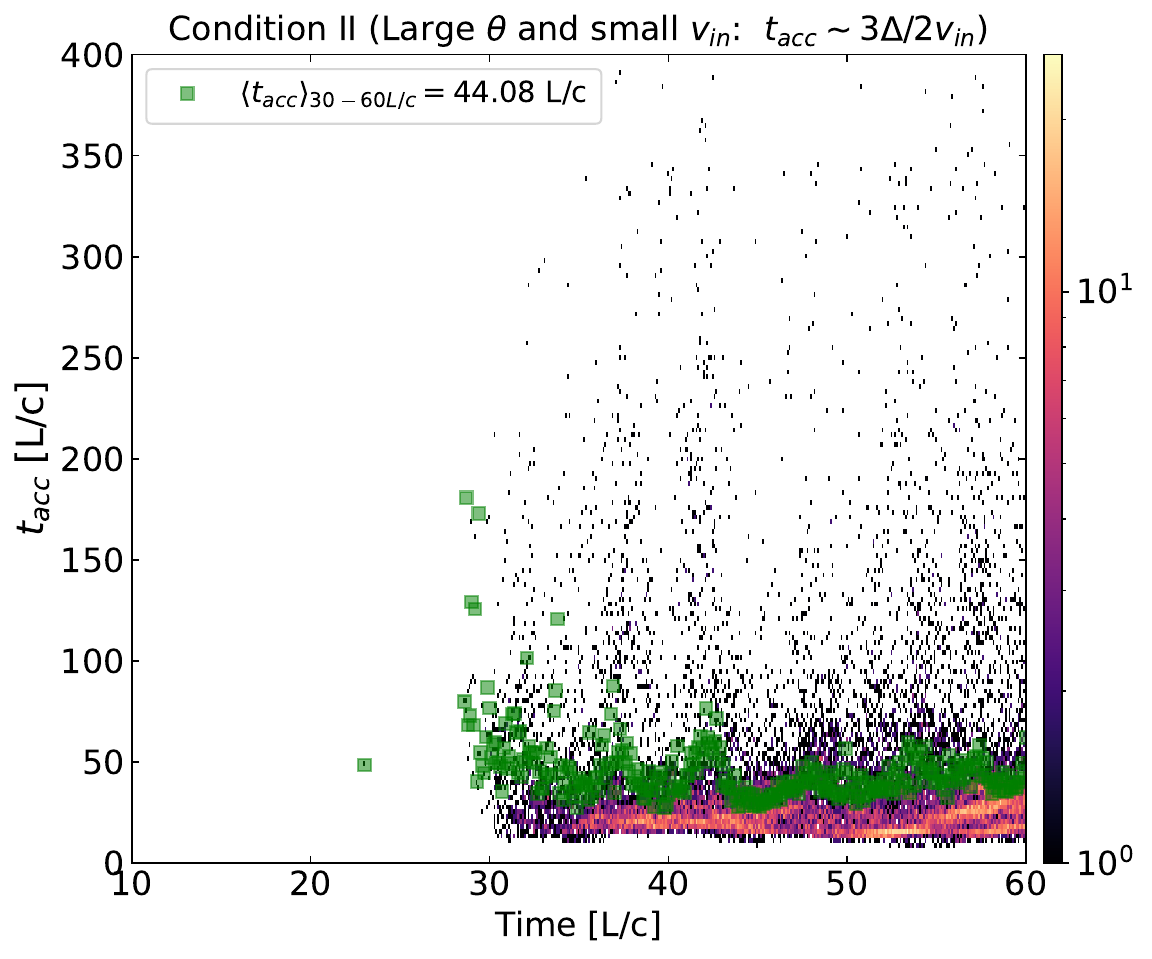}
   \includegraphics[scale=0.4]{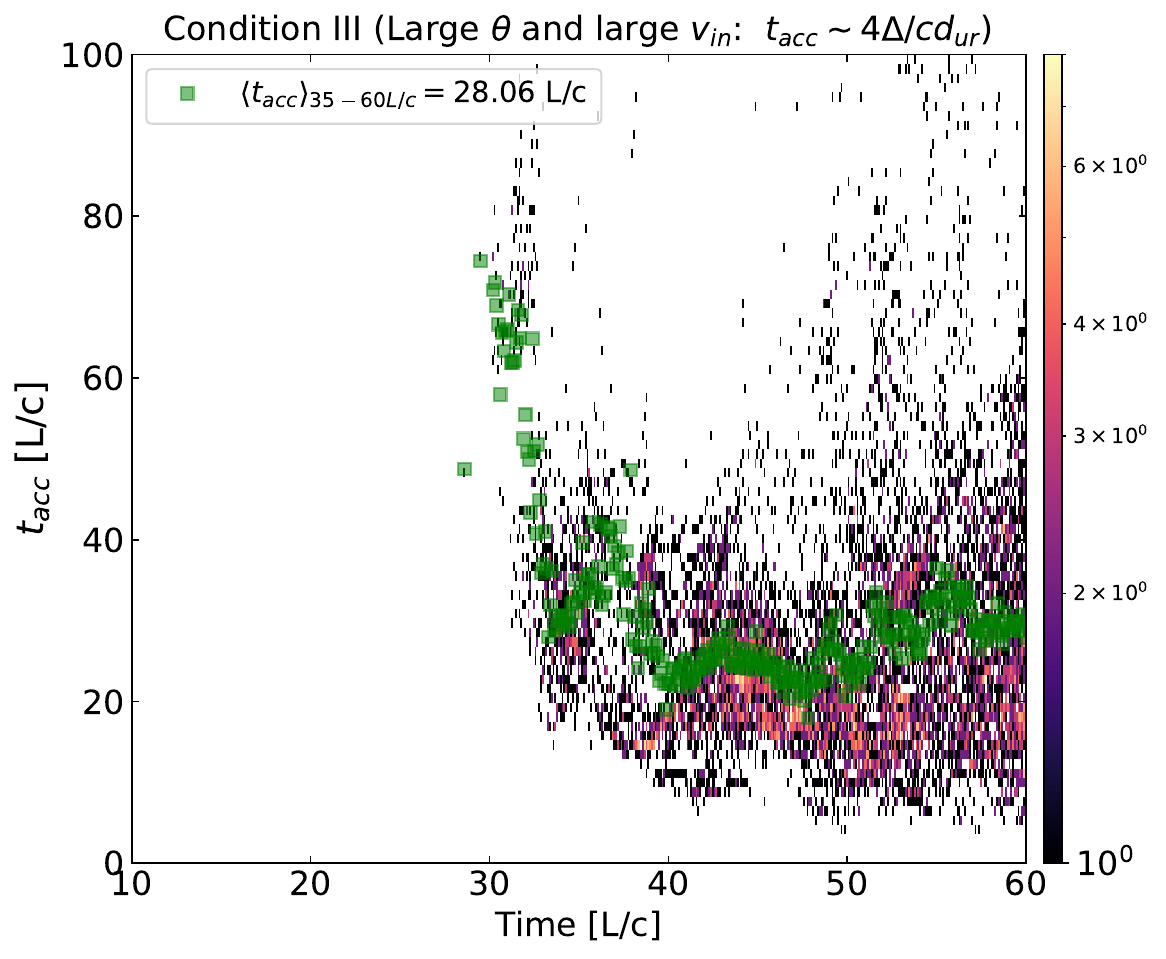}
   \caption{ Histograms of the acceleration time evolution, calculated from the magnetic reconnection sites identified by the search algorithm \citep{Kadowaki_2018,kadowaki_etal_2021}, using eqs. \ref{tacc1} to \ref{tacc3}. 
   Top: condtion I given by eq. \ref{tacc1}.
   Middle: condition II given by eq. \ref{tacc2}. 
   Bottom: condition III given by eq. \ref{tacc3}.
   The green squares give the mean acceleration time in each snapshot. 
   }\label{tacc123}
\end{figure}

\begin{figure}
 \centering
 \includegraphics[scale=0.5]{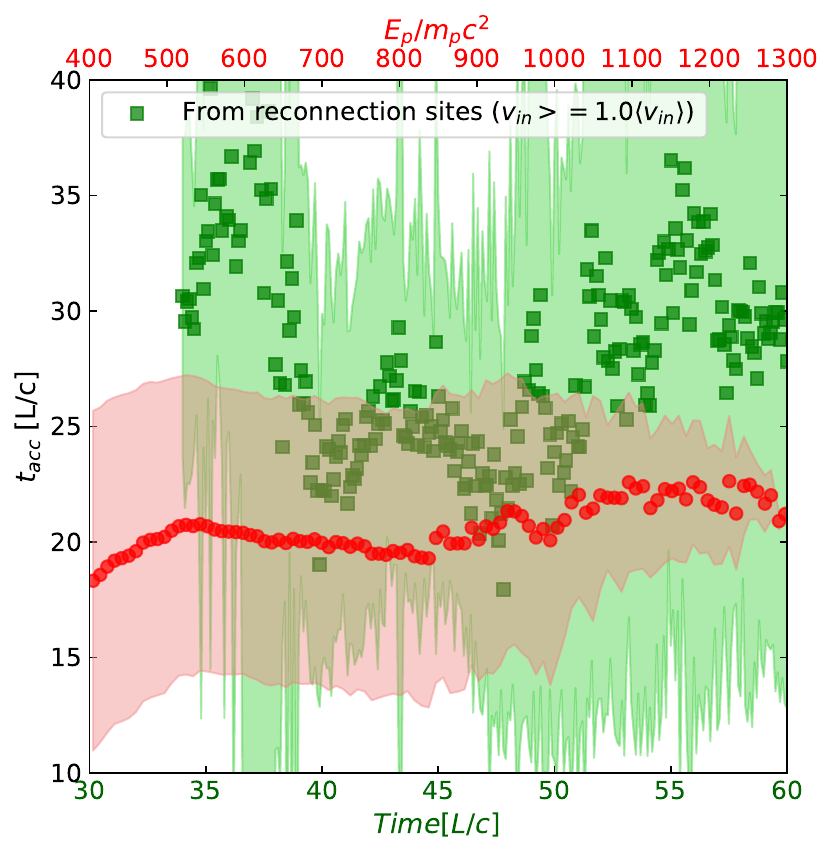}
   \includegraphics[scale=0.5]{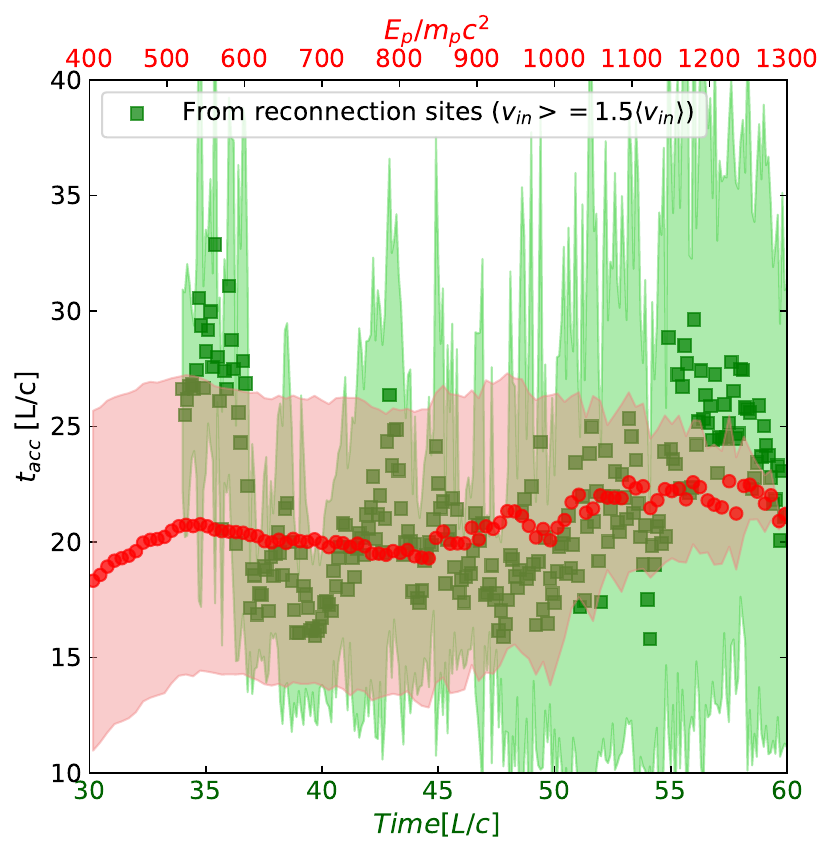}
\caption{  
Top panel: Average particle acceleration time calculated from two independent methods.  In red: the acceleration time as a function of both the particle kinetic energy normalized by the proton rest mass energy and the time, calculated directly from the particles accelerated in the MHD-PIC simulation of the 
jet, between $t=30$ and $60L/c$. In green: the acceleration time evolution, calculated directly from the fastest reconnection regions identified in the  jet from the condition III (eq. \ref{tacc3}), as in Figure \ref{tacc123} bottom. 
Bottom panel: the same as in the top diagram except that now the acceleration time in green, derived from condition III, is evaluated for values of reconnection velocity $v_{in} \geq 1.5 \left \langle v_{in}  \right \rangle$ in eq. \ref{tacc3} (instead of as in eq. \ref{array}).    The shaded regions correspond to the standard deviations. 
} 
\label{tacc-figure}
\end{figure}

Figure \ref{tacc123} shows  histograms of the acceleration time, $t_{acc}$, as a function of the evolution of the system, calculated for all identified reconnection sites in the 
jet for the three conditions  (eqs. \ref{tacc1}, \ref{tacc2} and \ref{tacc3}) in units of L/c. The distribution evolution of the corresponding parameters $\Delta$, $\theta$,  $\sin\theta$, and $v_{\rm rec}$, the reconnection velocity normalized by local Alfvén velocity
is given in Appendix \ref{appendix:rec_rate}. 

The green line (formed by square symbols) in Figure  \ref{tacc123} corresponds to the average acceleration time evolution.  
Before $t=30 L/c$, i.e., before the development of turbulence induced by the CDKI, there are almost no reconnection layers. The exception are a very few events detected in the top panel. The associated acceleration time, as inferred from condition I in this panel, is very large due to the very small reconnection velocities in such epochs (Figure \ref{hist-delta-theta-figure}; see also  Figure 4 in \citetalias{Medina-Torrejon_2023}).  
Interestingly, they occur   
around $t \sim 21-25 L/c$, where acceleration is mainly due to curvature drift in the wiggling magnetic field at the jet spine 
\citepalias[see][]{Medina-Torrejon_2023}.
We see in this regime the longest acceleration time values especially below t=30 L/c.
After t= 30 L/c the average acceleration time is $t_{acc} \sim 6700$ L/c. In condition II (middle panel), there is a decrease of the acceleration time, as expected, to an average $t_{acc}\sim 44$ L/c between dynamical  t= 30 to 60 L/c.
We find that most of the fast reconnection events lie in the condition III (bottom panel), having large $\theta$ and large $v_{in}$, with an average acceleration time  $t_{acc} < 30$ L/c.  We expect that particle acceleration will be predominantly influenced by these reconnection sites.

In Figure \ref{tacc-figure}, top diagram, we compare the acceleration time $t_{acc}$ calculated from the reconnection sites with the condition III, given by the green squares (the same as in  Figure \ref{tacc123}, bottom panel), with that calculated directly from the particles accelerated during the jet dynamical times  between $t \simeq 30 L/c$ and 60 $L/c$. Within this time interval, particles undergo exponential acceleration in the turbulent jet \citepalias[figure 6 top in ][]{Medina-Torrejon_2023}. The average values of the acceleration time are given by the red circles in this case. The shaded regions correspond to the standard deviations from the average of each evaluation. The acceleration time of the particles, 
$t_{acc} \simeq   \left \langle E_p  \right \rangle/  \left \langle \Delta E_p/\Delta t  \right \rangle$,
was calculated as in \citetalias{Medina-Torrejon_2023}, based on the average time per energy interval that particles take to reach a certain energy (see also \citetalias{Medina-Torrejon_2021} for more details). We began measuring the acceleration time of the test particles from when reconnection regions start to form in the relativistic jet, at $t =30 L/c$. We see that in condition III, the average value 
is consistent with
the average acceleration time calculated from the test particles,  $\sim 20.3 L/c$. The bottom diagram, shows the same comparison, but now, considering only the fastest reconnection events in condition III,
namely, those with $v_{in} \geq 1.5 \left \langle v_{in}  \right \rangle$ (instead of $v_{in} \geq 1.0 \left \langle v_{in}  \right \rangle$, as in eq. \ref{array}). In this case the average value given by the green curve is $\sim 21.3L/c$. While consistent with the average obtained in the upper diagram, it clearly aligns quite well with the 
the average acceleration time calculated from the test particles. The two acceleration time evaluations were conducted independently from each other and evidence an excellent agreement between the theoretical prediction (eq. \ref{tacc3}) and that  derived directly from the numerical simulation of the acceleration of test particles in the magnetic reconnection layers driven by the turbulence in the jet. 

The results above were obtained from the RMHD-PIC simulation where particles were accelerated during a small time interval within the evolving jet. For the physical parameters employed in the simulation $L \sim 5.2 \times 10^{-7}$ pc,
the total time elapsed in the jet evolution and the acceleration of the particles corresponds to $t_{total} = 60 L/c \simeq 1$  hr only \citepalias[see Table \ref{tablepartic} and also ][]{Medina-Torrejon_2023}.

\begin{figure}
 \centering
   \includegraphics[scale=0.4]{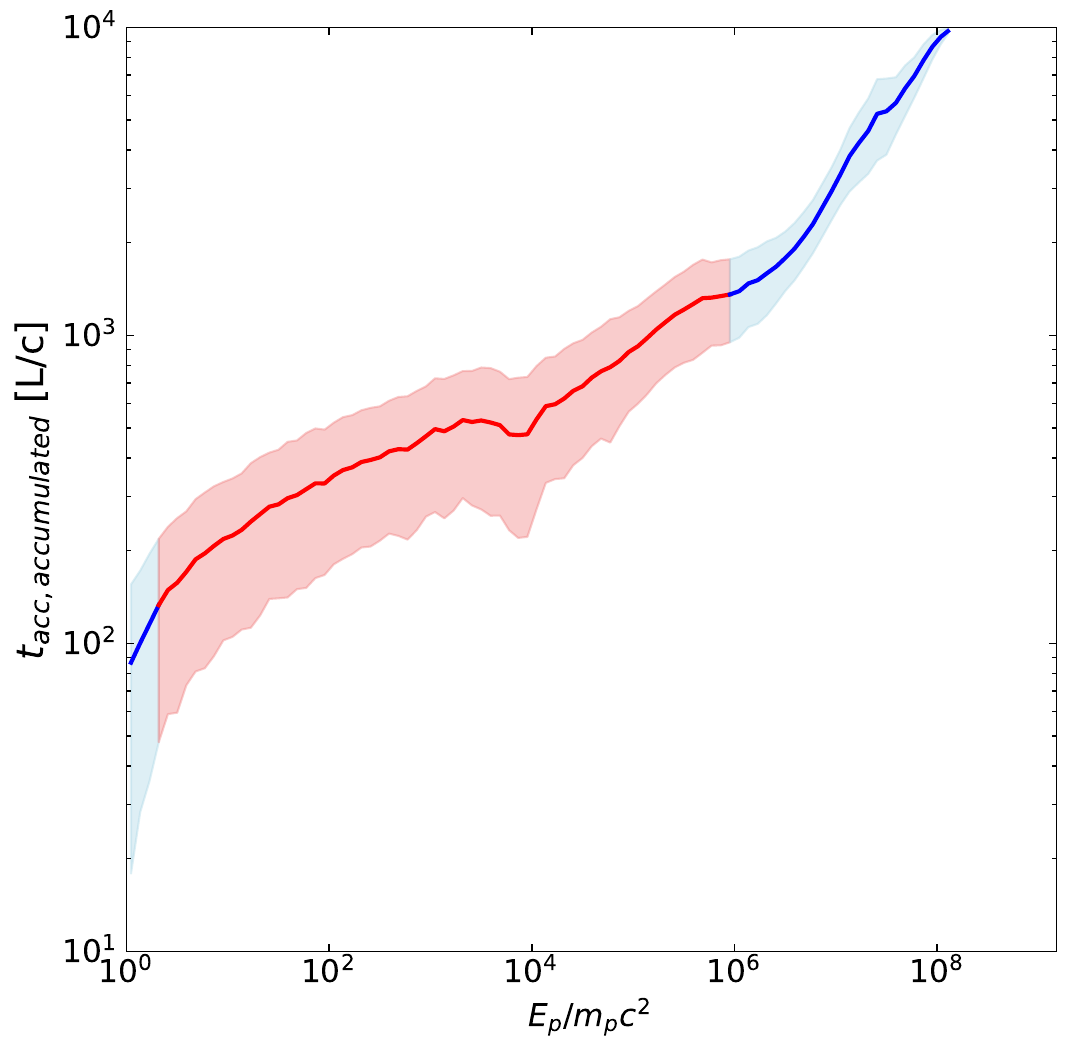}
   \includegraphics[scale=0.4]{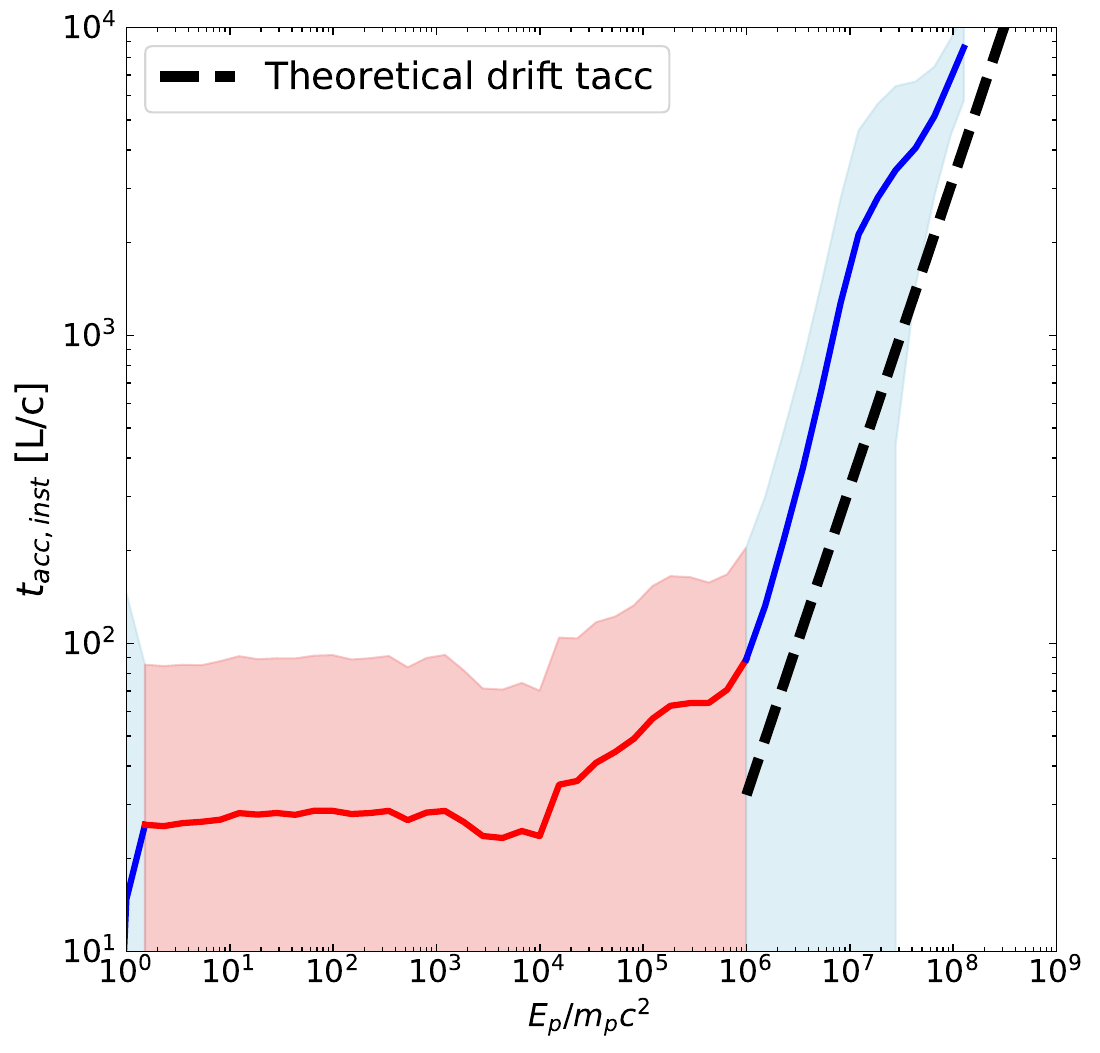}
\caption{  
Top panel: accumulated acceleration time of the particles injected in the nearly stationary turbulent jet at snapshot 
t= 45 L/c 
(RMHD-GACCEL model in Table \ref{tablepartic})
as a function of the energy. The derivation is the same as in \citetalias{Medina-Torrejon_2021} (see their Figure 9),  where the acceleration time is evaluated considering the total particles energy increase from the beginning until each energy.  Bottom panel: displays the instantaneous acceleration time obtained directly  from $t_{acc, inst} = E_p/(dE_p/dt)$ (see also bottom panel of Figure 9 in \citetalias{Medina-Torrejon_2021}). The acceleration time in both diagrams is given in units of L/c. For this model, the physical value of L is 
$3.5 \times 10^{-5}$ pc.  
The red color characterizes the regime of particle acceleration with approximately exponential energy growth of their energy, where the acceleration time is approximately constant (i.e. the Fermi regime within reconnection layers), while the blue color characterizes the regime of drift acceleration in the spatially variable non-reconnected magnetic fields, both in the very begginning of particle injection and later on, after the accelerated particles scape the reconnection layers. The dashed line shows the theoretical prediction for the  acceleration time in the drift regime which aligns very well with the numerical curve. The shaded regions give the average statistical errors. 
}
\label{taccfit-figure}
\end{figure}

We can compare the particle acceleration time $t_{acc}$ obtained in this small time interval, which reflects a nearly  instantaneous acceleration, with that obtained from  test particles accelerated in one of the nearly stationary fully turbulent snapshots of the simulated jet  for a much longer time. Figure \ref{taccfit-figure} depicts diagrams of the acceleration time in this case, obtained from  test particles injected in the jet  at the turbulent snapshot 
$t=45$ L/c, 
using the  \texttt{GACCEL} code (model RMHD-GACCEL in Table \ref{tablepartic}; see also  figure 5 in \citetalias{Medina-Torrejon_2023}). The top diagram shows the "accumulated" acceleration time as a function of the particles energy. This is calculated from the accelerated particles as in 
\citetalias{Medina-Torrejon_2021} and \citetalias{Medina-Torrejon_2023},
based on the average time per energy interval that particles take to reach a certain energy. However, this evaluation considers the total particle energy increase from the beginning until each energy. The bottom panel of the same figure, on the other hand, provides the "instantaneous" acceleration time, i.e., $t_{acc, inst} = E/(dE/dt)$, as a function of time. We see that this is comparable to the one obtained in the short time interval of Figure  \ref{tacc-figure}.  Figure  \ref{tacc-figure} actually corresponds to the tiny portion around $E \simeq (10^2 - 10^3) m_p c^2 $ in the bottom panel of Figure \ref{taccfit-figure}.

Both diagrams in Figure \ref{taccfit-figure} show similar  shape and trend, starting with a regime  where  $t_{acc}$ is weakly dependent on particle energy\footnote{The dependence of \( t_{acc} \) on \( E_p \) for this model was  analyzed in \citepalias[][]{Medina-Torrejon_2021, Medina-Torrejon_2023}. On average, across the entire red zone, it follows a relation of \( t_{acc} \sim E_p^{0.1} \) (see Figure 8 of \citetalias{Medina-Torrejon_2023}).},
corresponding to Fermi acceleration within the reconnection layers (red zone). 
In this phase, kinetic energy grows approximately exponentially over time (see  Figure 5 in \citetalias{Medina-Torrejon_2023}).
This is followed by a drift regime of acceleration (blue zone) beyond the threshold energy of the Fermi regime, where $t_{acc}$ exhibits a much stronger dependence on the energy. The transition to the drift regime occurs when the particles' Larmor radius exceeds the maximum thickness of the current sheets \citep[e.g.][]{kowal_etal_2012, delvalle_etal_16, Zhang2023}, \citepalias[][]{Medina-Torrejon_2021, Medina-Torrejon_2023}.
This maximum thickness is of the order of the injection scale of the turbulence,  
which does not exceed the jet diameter \citepalias[][]{Medina-Torrejon_2021, Medina-Torrejon_2023}. According to  Figure \ref{taccfit-figure} particles leaving the Fermi regime in the RMHD-GACCEL model have  $E_p \sim 10^6 m_p c^2$. For the average magnetic field of the simulation, $B \sim 0.1$ G \citepalias[][]{Medina-Torrejon_2021, Medina-Torrejon_2023}, the corresponding particle  Larmor radius is  $r_L \simeq 3.3 \times 10^{13}$ cm $\sim 1 L$ (see Table \ref{tablepartic}). This is consistent with the maximum thickness of the reconnection layers depicted in Figure \ref{hist-delta-theta-figure} (left) in Appendix \ref{appendix:rec_rate}, which  indicates $\Delta_{max} \sim L$.

Looking at the bottom panel of Figure \ref{taccfit-figure} in more detail, we see that until $E_p \simeq 2 \times 10^4$ $m_p c^2$, the average numerical value of  $t_{acc}$ matches nicely with the theoretical prediction of fastest reconnection given by condition III (eq. \ref{tacc3}, bottom panel of Figure \ref{tacc123}, and Figure \ref{tacc-figure}), while between $E_p \simeq 2 \times 10^4$ $m_p c^2$ and $\simeq 10^6$ $m_p c^2$,  there is slight increase in  average $t_{acc}$ which 
matches better 
with  the condition II (eq. \ref{tacc2}, and middle panel of  Figure \ref{tacc123}), and finally with condition I (eq. \ref{tacc1}, and top panel of  Figure \ref{tacc123}), indicating  slight changes in the reconnection rate to intermediate and smaller values. Both intervals are still in the Fermi regime of particle acceleration within the reconnection sites. 
In other words, particles are taking progressively longer to accelerate (indicating smaller reconnection velocities in the associated reconnection layers), signaling a smooth gradual transition to the drift regime. 
This is connected to the fact that the reconnection layers capable of confining more energetic particles are also larger, given the larger Larmor radius of these particles.
This smooth transition from conditions III and  II to I may be also reflecting the effects of decreasing obliquiness—that is, a smaller angle $\theta$ between the reconnecting magnetic field and the guide field—which leads to an increase in $t_{\text{acc}}$ due to the influence of a stronger guide field, as predicted by Eq. \ref{tacc1}.


The dashed line in the bottom diagram of Figure \ref{taccfit-figure} shows the theoretical approximation for the acceleration time in the drift regime (eq. \ref{adrift}).
It was calculated employing the average values in the simulation $B\simeq 0.1$  G and $v_{\rm rec} \simeq 0.03 v_A \simeq 0.01 c$.
We note that there is a good agreement between the theoretical prediction and the numerical simulation also for this regime of acceleration.

\section{Discussion and Conclusions} \label{sec:discussion}

In this work we have investigated numerically a crucial parameter for the comprehension of particle acceleration theory by turbulence-induced magnetic reconnection, namely,  the  particle acceleration time (or rate). 

Employing recent 3D numerical simulations of magnetically dominated relativistic jets with particles being accelerated either during the dynamical time evolution of the jet \citepalias[][]{Medina-Torrejon_2023},
 or accelerated in a post-processing, nearly stationary regime of the jet \citepalias[][]{Medina-Torrejon_2021, Medina-Torrejon_2023}, we have derived the particle acceleration time and compared with theoretical predictions for both acceleration regimes identified in the simulations, the Fermi regime of magnetic reconnection acceleration and the drift regime which follows.
 
In the Fermi regime, the acceleration time is expected to be nearly independent of the particles energy, for a constant reconnection 
velocity, since this energy increases exponentially with time 
\citep[][]{dalpino_lazarian_2005, drake_etal_2006,  kowal_etal_2012, dalpino_kowal_15, guo_etal_2019}, \citepalias[][]{Medina-Torrejon_2021, Medina-Torrejon_2023}.
On the other hand, we expect a  dependence of the reconnection acceleration time with the thickness of the current sheet and the reconnection velocity. This dependence has been recently revisited by \citetalias[][]{xu_lazarian_2023}. They identified three different conditions for $t_{acc}$ within the reconnection layer, depending on the reconnection velocity and the angle between the reconnecting field and the background guide field.   We tested their relations  using the statistical distributions of the angle, thickness, and reconnection velocities of the current sheets detected in the turbulent jet along time. The resulting average value of $t_{acc}$ was found to be nearly constant with the particles energy, as expected. 

Moreover, this acceleration time has been compared with the  acceleration time evaluated directly from the    50,000 particles  accelerated in situ in the same relativistic jet (Figure \ref{tacc-figure}).  
Our results have evidenced an excellent  agreement between these two independent evaluations, especially for  the fastest reconnection acceleration condition (III) given by eq. \ref{tacc3}. 

The analysis in Figure \ref{tacc-figure}  was performed with particles being accelerated while the relativistic jet was dynamically evolving (in a RMHD-PIC simulation, Table \ref{tablepartic}). This allowed us to compare both derivations of the acceleration time, 
but in a small time and energy interval.

When we take the much larger time interval for particles acceleration injecting them in a nearly stationary snapshot of the same turbulent jet (RMHD-GACCEL model in Table \ref{tablepartic}), we find that the instantaneous acceleration time during the Fermi regime is nearly independent of the particles energy and also comparable to the acceleration time relations derived by 
\citetalias{xu_lazarian_2023} until the threshold energy $E_{th} \simeq 10^6$  $m_p c^2 $. Beyond this, the acceleration regime changes to the slower drift regime with  strong dependence on the particles energy (in agreement with eq. \ref{adrift}), resulting much longer acceleration times.

The results above demonstrate the robustness and consistency between both the theoretical predictions and the numerical simulations, particularly regarding the Fermi mechanism operating within the magnetic reconnection layers up to a threshold energy. This threshold is reached when the Larmor radius of the particles reach the thickness of the largest reconnection layers, which is determined by the injection scale of the turbulence. In the jet, this scale is on the order of $L$ (Figure \ref{hist-delta-theta-figure}
in Appendix \ref{appendix:rec_rate}) and does not exceed the jet diameter \citepalias[][]{Medina-Torrejon_2021, Medina-Torrejon_2023}.
Similar results have been observed previously in 3D MHD simulations of single turbulent current sheets \citep[e.g.][]{kowal_etal_2012, delvalle_etal_16}.
 Beyond these scales, particles undergo further acceleration by drift in non-reconnected fields, albeit at a slower rate. Since particles achieve very high energies within the Fermi regime and continue to gain energy, albeit more slowly, in the drift regime, both regimes are crucial in shaping the particle spectrum at high and very high energies, with the Fermi regime being dominant. This is supported by the 3D MHD large-scale simulations presented above where particles are accelerated to very high and ultra-high energies within the Fermi regime \citepalias[][]{Medina-Torrejon_2021, Medina-Torrejon_2023}.
 This contrasts with recent conclusions based on kinetic simulations that claim that the particle spectrum is predominantly shaped by drift acceleration \citep{zhang2021, Zhang2023}, highlighting the need for caution when extrapolating kinetic results to very large macroscopic scales. Furthermore, considering the non-thermal radiative losses that particles may undergo within systems like blazars, they might lose a substantial part of their energy even before reaching the threshold of the Fermi regime and thus, the transition to the drift regime \citep[e.g.][]{dalpino2024}.


 Although radiative losses are not considered in this work, we can still compare the acceleration time with other relevant timescales, such as the particle escape time from the system and the scattering diffusion time.

 First of all, it is important to note that in the Fermi acceleration regime inside  the reconnection layers,
scattering diffusion is not essential for 
confinement. 
Particles gain energy primarely through head-on interactions with inflows generated by magnetic reconnection, and confinement  is primarily due to magnetic mirroring and the convergence of magnetic field lines.
Particles can move ballistically along the magnetic field lines 
when pitch-angle scattering by turbulence is inefficient.
This ballistic motion leads to much shorter intervals between successive acceleration events compared, for instance, to the slower diffusive motion seen in diffusive shock acceleration \citep[see detailed discussion e.g. in][]{xu_lazarian_2023} 
 \citep[see also][]{Barreto-Mota2024}. 
Consequently, while particles remain in the Fermi regime and are accelerated within reconnection layers, scattering diffusion is not a critical factor. 
On the other hand, in the less efficient drift regime of acceleration that follows, diffusive effects become more significant, allowing us to analyze the diffusion time characteristic of this regime.

In the RMHD-GACCEL simulation of Figure \ref{taccfit-figure}, particles are able to  “re-enter” the jet several times along the longitudinal z-direction which has periodic boundaries. This allows them to reach the threshold energy of the Fermi regime and then undergo further acceleration by drift in the non-reconnected fields \citepalias[][]{Medina-Torrejon_2021, Medina-Torrejon_2023}.
To compare the different implied timescales, we refer to the diagram of kinetic energy growth versus time from this test particle simulation  (see Figure 5 of \citetalias{Medina-Torrejon_2023}). 
This diagram
indicates that particles reaching
an energy of  $\sim 10^7 mc^2$ are in the drift-dominated phase.
This  diagram also shows that particles with this energy have evolved  for  $\gtrsim 5000$ hr $= 5000 L/c$, and since particles travel at relativistic speeds ($v_p \simeq c$), the corresponding effective physical path length they travelled along the jet is $L_j \gtrsim 5000 L$ \footnote{According to the scale relations of the test particle simulation RMHD-GACCEL,  $L = 3.5 \times 10^{-5}$ pc (Table \ref{tablepartic}), giving $L_j \gtrsim 5000 \, L \simeq  0.2$  pc. This path length does not imply that the jet itself has this physical size. Rather, it reflects the effective distance particles travel within the acceleration region standard approach in test-particle studies of extended systems. This effective extension $\sim 0.2$ pc is, in fact, consistent with the expected scale of the inner regions of AGN blazar jets, near the central engine where the plasma remains magnetically dominated and reconnection-driven acceleration is expected to prevail \citepalias[e.g.][]{Medina-Torrejon_2021}.}.
 This implies an escape time $t_{esc} \gtrsim 5000L/c$.  This escape time can be compared with the acceleration time  $t_{acc}$ for the same particles, as shown in Figure \ref{taccfit-figure} (bottom) of this work. We find $t_{acc} \sim1000L/c$, which is significantly shorter than $t_{esc}$. 
For comparison with the scattering diffusion time in this drift regime, we use a simple estimate based on Bohm-like diffusion, which gives \( t_{\text{diff}} \sim \frac{3}{2} \frac{d^2}{c r_L} \),

where $d$ is the longitudinal diffusion length scale and $r_L$  is the Larmor radius of the particle  with energy $E_p$ in the background field $B \sim 0.1$  G (in the GACCEL-RMHD simulation, \citetalias{Medina-Torrejon_2021, Medina-Torrejon_2023}).
For example, for particles with \( E_p \sim 10^7 mc^2 \), 
\( t_{\text{diff}} \) is comparable to \( t_{\text{acc}} \) for a longitudinal diffusion scale is \( d \sim 46L \).  This diffusion scale is much smaller than the effective  jet  scale length  travelled by the particles ($L_j$) by the time they reach this energy, implying an efficient confinement. These results indicate that, in the absence of radiative losses, particles can be accelerated to such energies before escaping the system, even during the slower drift acceleration regime.

Finally, we should notice that   a detailed study of the resulting  
 energy spectrum of the particles ($ \propto E_p^{-\alpha}$) has already been conducted in \citetalias{Medina-Torrejon_2023}  for the RMHD-PIC simulation and in \citepalias[][]{Medina-Torrejon_2021}  for the RMHD-GACCEL simulation.
In \cite{xu_lazarian_2023} work, they predict  a range of spectral indices, $\alpha \simeq 2.5–4$ for particles being accelerated within the reconnectin layers, with the steepest spectra ($\alpha \simeq 4$) occurring under conditions of strong guide fields—i.e., small angle $\theta$ between the inflow magnetic field and the guide field. In their analysis, this upper limit is reached when both the inflow velocity $v_{in}$   and $\theta$ are small. As $\theta$ increases (weaker guide field), $\alpha$ tends to decrease due to higher particle energy gains at larger obliqueness. 
The other threshold condition they identified, at $\alpha \simeq 2.5$  \citep[see also][]{dalpino_lazarian_2005}, applies to cases of large $\theta$ (weak guide field), but  small $v_{in}$  (see their Eq. 62). Since their model does not explore the full range of realistic $v_{in}$  values, it does not offer a precise prediction for $\alpha$ when both $\theta$ and $v_{in}$ are large, but their numerical results indicate that for  $very$ large velocities $\alpha \simeq 3.3-3.5$ for large $\theta$.  On the other hand, both PIC and MHD simulations have shown that under large enough velocities and large $\theta$ conditions,
flatter spectral indices can emerge, consistent with efficient Fermi acceleration within the reconnection layer. For example, in Figure 7 (top) of \citepalias[][]{Medina-Torrejon_2023}, the spectrum displays two distinct tails: a flatter one at lower energies and a steeper one at higher energies, both evolving over time. Once turbulence becomes fully developed in the jet around t=35 c.u., the low-energy tail follows a power-law distribution with an index $\alpha$ varying from approximately 2.9 to 0.8 between t=35 and t=51 c.u. Meanwhile, the high-energy tail steepens, with $\alpha$ evolving from about 3.4 to 4.8 between t=40 and t=51 c.u. The flatter spectral index observed in the initial portion of the spectrum at t=51 c.u. is consistent with a Fermi acceleration regime, as reported by \citepalias[][]{Medina-Torrejon_2023}, and aligns with earlier findings from both PIC and MHD studies. However, this value does not represent the full high-energy asymptotic behavior, which may be influenced both by the obliqueness $\theta$, as discussed in \cite{xu_lazarian_2023}, and by the subsequent drift-dominated regime that emerges once particles exit the current layers—beyond the Fermi phase. This latter regime, while not considered in \cite{xu_lazarian_2023},  is clearly evident in our simulations \citepalias[][]{Medina-Torrejon_2021, Medina-Torrejon_2023}. We refer readers to these works for a more detailed analysis. However, we did not explore the dependence on $\theta$ in those studies, and we plan to address this analysis in future publication.


\begin{acknowledgments}
 EMdGDP and TEMT acknowledge support from the Brazilian Funding Agency FAPESP (grants 2013/10559-5,   2021/02120-0, 2023/08554-7).  EMdGDP also acknowledges support from CNPq (grant 308643/2017-8).
This research was also supported in part by grant no. NSF PHY-2309135 to the Kavli Institute for Theoretical Physics (KITP), and EMdGDP acknowledges fruitful discussions with the participants of the program on “Turbulence in Astrophysical Environments” at KITP. The authors also wish to express their gratitude to Greg Kowal, Alex Lazarian, and Syao Shu for invaluable discussions. 
The simulations presented in this work were performed using the cluster of the Group of Plasmas and High-Energy Astrophysics at IAG-USP (GAPAE), acquired with support from FAPESP (grants 2013/10559-5 and 2021/02120-0).
\end{acknowledgments}

\section*{Data Availability}
The data of this article will be available upon request to the  authors.

%



\software{ VisIt \citep{HPV:VisIt},  
          Python3 \citep{python3}
          }.

\newpage

    
\appendix

\section{Histograms of the Parameters of the Reconnection Sites identified in the Simulated Jet} \label{appendix:rec_rate}


\begin{figure}
    \centering
    \includegraphics[width=0.49\linewidth]{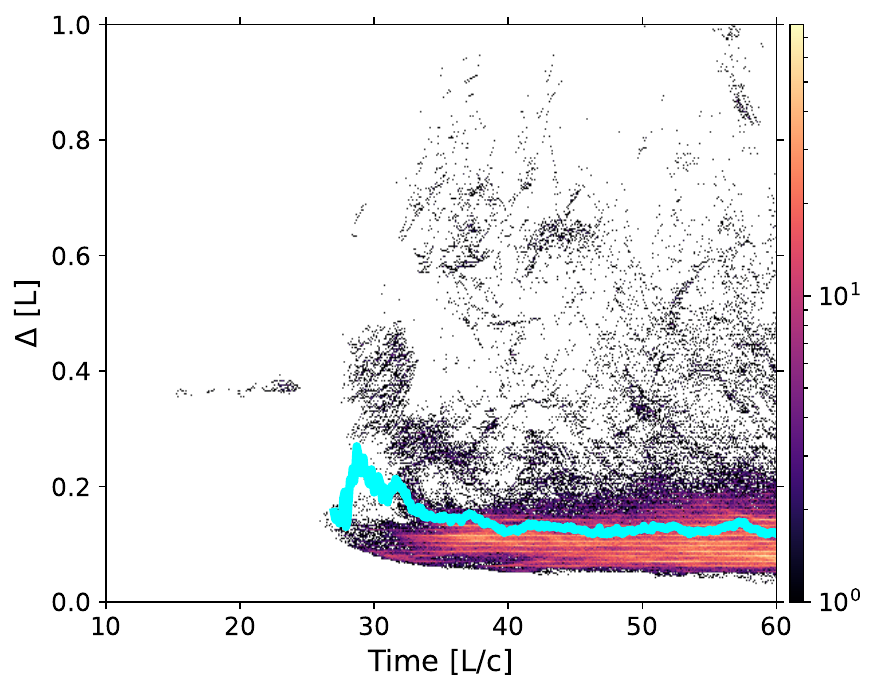}
    \includegraphics[width=0.49\linewidth]{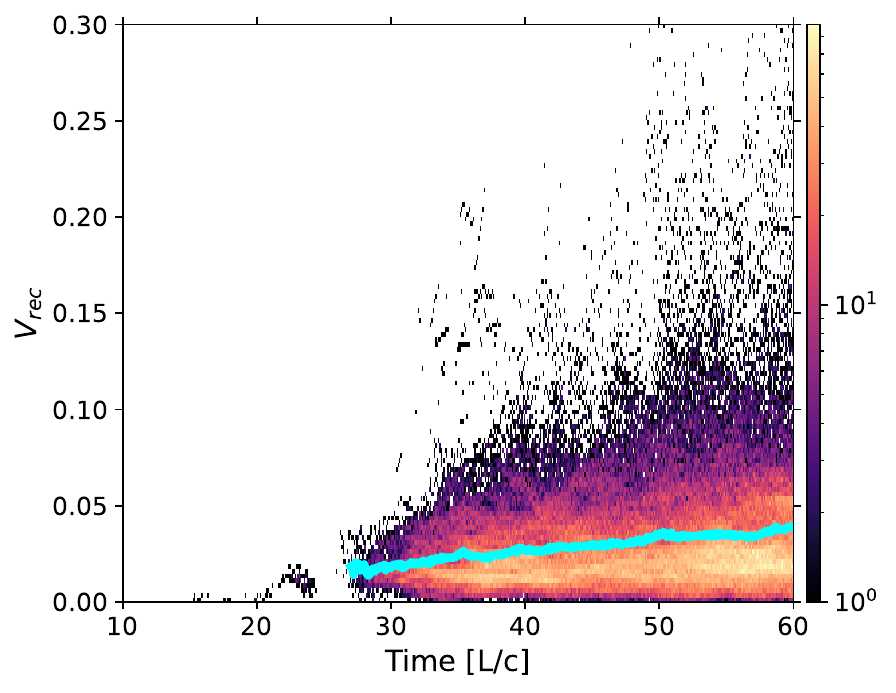}
    \includegraphics[width=0.49\linewidth]{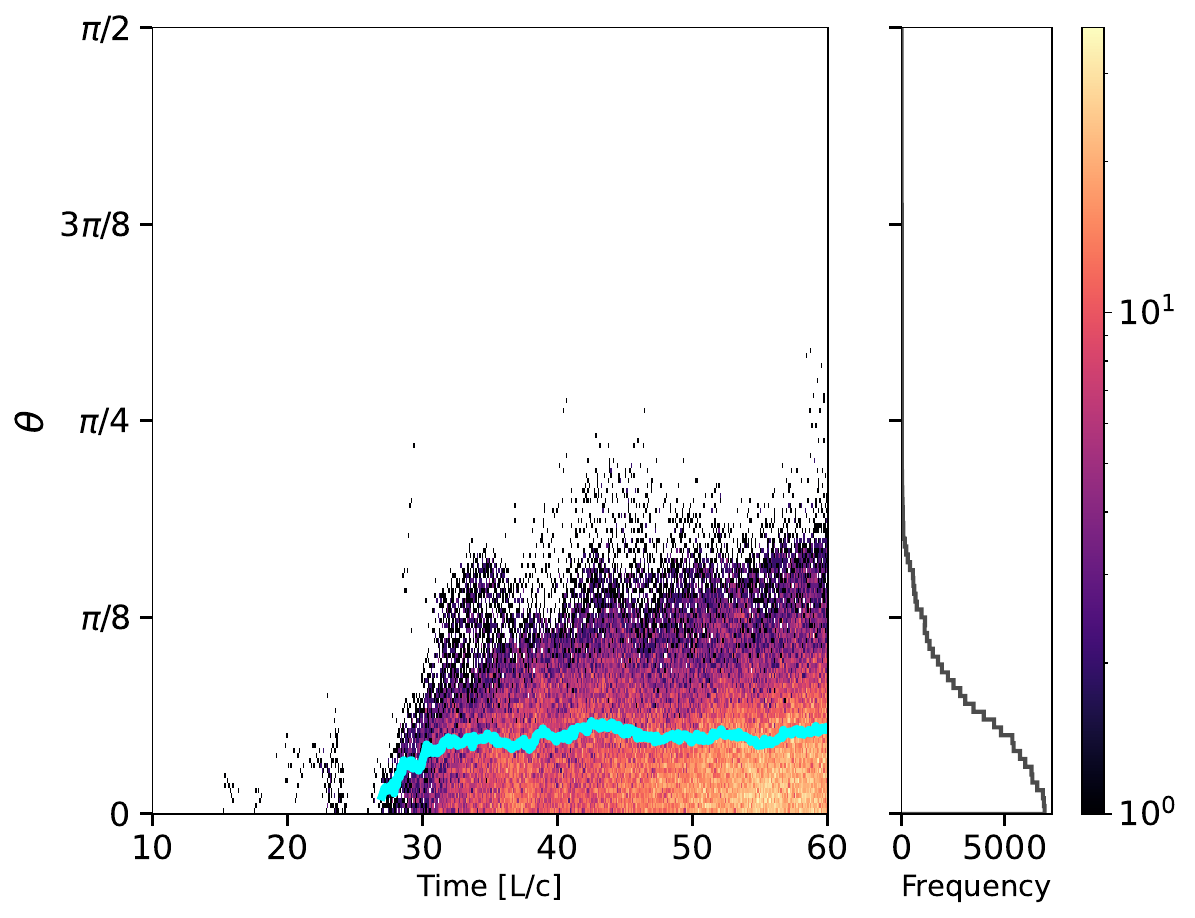}
    \includegraphics[width=0.49\linewidth]{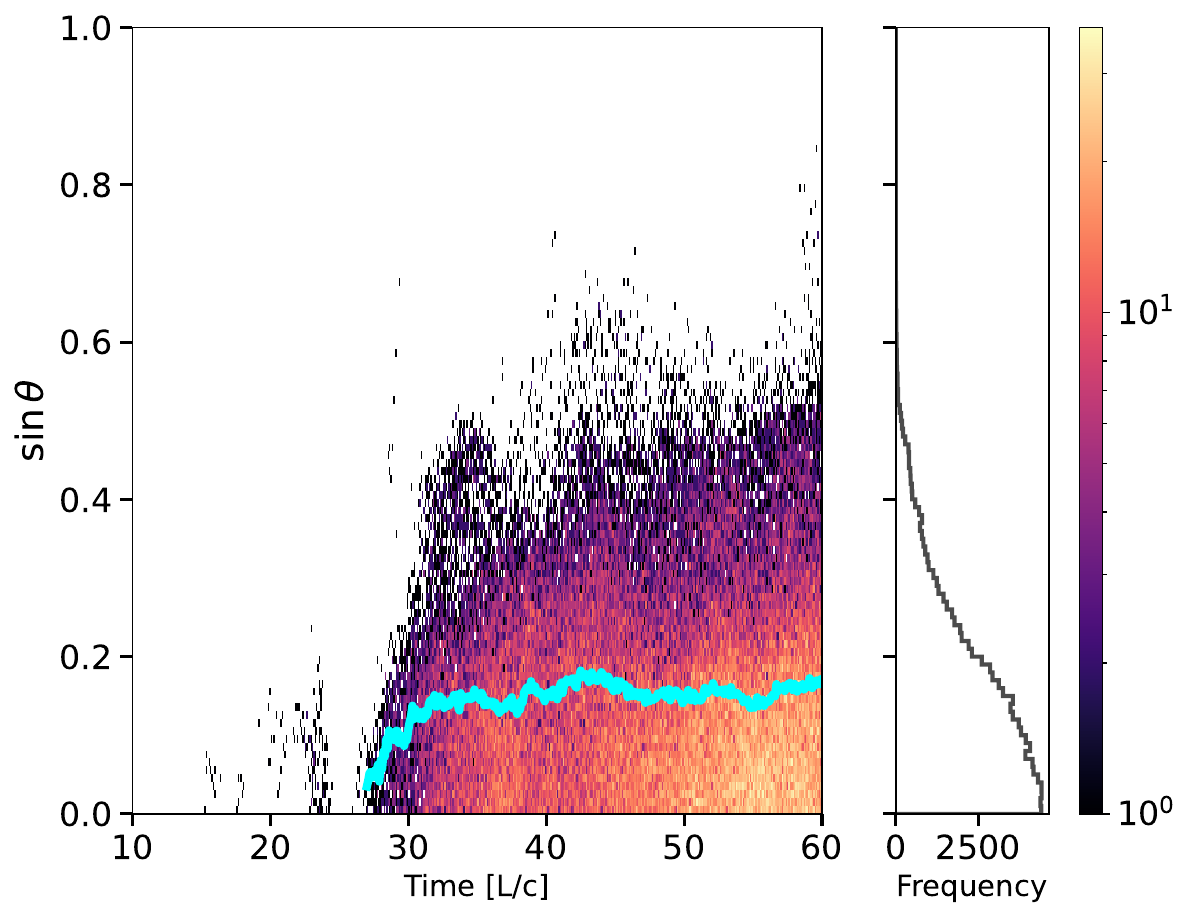}
    \caption{Temporal evolution of the magnetic reconnection parameters. The color maps display the
    histograms as functions of the simulated jet dynamical time for: the current sheet thickness $\Delta$ (top left), the reconnection velocity $v_{\rm rec}$ in units of the Alfven speed (top right), the  angle $\theta$ (bottom left), and the  $\sin \theta$ (bottom right). The  cyan lines indicate the mean values at each time step. These are plotted starting from $t = 27 $ L/c to focus on the evolved stage of turbulent-driven reconnection. For the bottom panels, marginal distributions integrated over the entire temporal domain are also shown as black-stepped histograms on the right-hand side of each plot.}
    \label{hist-delta-theta-figure}
\end{figure}

Figure \ref{hist-delta-theta-figure} shows histograms of the current-sheet thickness $\Delta$, the reconnection velocity $v_{in} = v_{\rm rec}$,  the angle $\theta$ between the inflow magnetic field to the reconnection layer and the guide field, and  $\sin \theta$, as  functions of the dynamical evolution of the simulated jet. 
These quantities are used to derive the acceleration time evolution from eqs. \ref{tacc1}-\ref{tacc3}, shown by the green curves in Figures \ref{tacc123} and   \ref{tacc-figure}.

They were computed  
using the reconnection-site search algorithm detailed in \citet[][]{kadowaki_etal_2021} \citep[see also][]{Kadowaki_2018}. The  distribution of reconnection velocities shown in the top right panel is the same as that presented in \citetalias{Medina-Torrejon_2023}.

We note that, on average, the thickness of the reconnection layers is found to be $\Delta \simeq 0.1 L$, though values as large as $\sim 1L$ can also occur.
The angle $\theta$ 
tends to cluster below $\sim \pi/8$. Likewise, the corresponding  $\sin \theta$ distribution integrated over the entire domain  shows the dominance of $\sin \theta < 0.5$, and an average  $\sim 0.2$.  We employ this value in eq. \ref{array1} to define the separation between large and small values of $\sin \theta$.







\bibliography{sample631}{}
\bibliographystyle{aasjournal}




\end{document}